\documentclass[sigconf]{acmart}
\usepackage[utf8]{inputenc}
\usepackage{balance}
\usepackage{amsmath,amsfonts}
\usepackage{algorithm}
\usepackage[noend]{algpseudocode}
\usepackage{mdframed}
\usepackage{courier}
\usepackage{enumitem}
\usepackage{hyperref}
\usepackage{listings}
\usepackage[most]{tcolorbox}
\usepackage{mathtools}
\usepackage{xspace}
\usepackage{array,multirow}
\usepackage{xcolor}
\usepackage{colortbl}
\usepackage{mdframed}
\usepackage{graphicx}
\usepackage{caption}
\usepackage{subcaption}

\usepackage{listings}
\usepackage{xcolor}

\definecolor{pythonblue}{RGB}{0,0,255}
\definecolor{pythongreen}{RGB}{0,128,0}
\definecolor{pythonpurple}{RGB}{128,0,128}
\definecolor{pythongray}{RGB}{128,128,128}

 \definecolor{boxcolor}{RGB}{238, 223, 204} %
\newcommand{\mybox}[2][gray!20]{%
\begin{tcolorbox}[   %% Adjust the following parameters at will.
        breakable,
        left=0pt,
        right=0pt,
        top=0pt,
        bottom=0pt,
        colback=#1,
        colframe=black,
        width=\dimexpr\columnwidth\relax, 
        enlarge left by=0mm,
        boxsep=5pt,
        outer arc=4pt,
        boxrule=.5mm
        ]
        #2
\end{tcolorbox}
}
\lstdefinestyle{python}{language=Python,
    language=Python,
    basicstyle=\footnotesize\ttfamily,
    numbers=left,
    numberstyle=\tiny,
    numbersep=5pt,
    showstringspaces=false,
    framesep=2mm,
    xleftmargin=1.8em,
    tabsize=4,
    breaklines=true,
    breakatwhitespace=true,
    keywordstyle=\color{pythonblue},
    commentstyle=\color{pythongreen},
    stringstyle=\color{pythonpurple},
    identifierstyle=\color{black},
    numberstyle=\color{pythongray},
    escapeinside={!}{!},
    morekeywords={assert, given}
}

\newcommand{\code}{\texttt}
 
\definecolor{clcolor}{rgb}{0.5,0.7,0.9}

\newcommand{\rqn}[2]{\noindent \textbf{RQ#1}: \emph{#2}}

\newcommand{\Chi}{\mathrm{X}}

\title{Can Large Language Models Write Good Property-Based Tests?}

\author{Vasudev Vikram}
\affiliation{%
  \institution{Carnegie Mellon University}
  \city{Pittsburgh, PA}
  \country{United States}}
\email{vasumv@cmu.edu}

\author{Caroline Lemieux}
\affiliation{%
  \institution{University of British Columbia}
  \city{Vancouver, BC}
  \country{Canada}}
\email{clemieux@cs.ubc.ca}

\author{Joshua Sunshine}
\affiliation{%
  \institution{Carnegie Mellon University}
  \city{Pittsburgh, PA}
  \country{United States}}
\email{sunshine@cs.cmu.edu}

\author{Rohan Padhye}
\affiliation{%
  \institution{Carnegie Mellon University}
  \city{Pittsburgh, PA}
  \country{United States}}
\email{rohanpadhye@cmu.edu}

\begin{document}
\sloppy

\begin{abstract}

Property-based testing (PBT), while an established technique in the software testing research community, is still relatively underused in real-world software. Pain points in writing property-based tests include implementing diverse random input generators and thinking of meaningful properties to test. Developers, however, are more amenable to writing documentation; plenty of library API documentation is available and can be used as natural language specifications for PBTs. As large language models (LLMs) have recently shown promise in a variety of coding tasks, we investigate using modern LLMs to automatically synthesize PBTs using two prompting techniques. A key challenge is to rigorously evaluate the LLM-synthesized PBTs. We propose a methodology to do so considering several properties of the generated tests: (1) validity, (2) soundness, and (3) \textit{property coverage}, a novel metric that measures the ability of the PBT to detect property violations through generation of property mutants. In our evaluation on 40 Python library API methods across three models (GPT-4, Gemini-1.5-Pro, Claude-3-Opus), we find that with the best model and prompting approach, a valid and sound PBT can be synthesized in 2.4 samples on average. We additionally find that our metric for determining soundness of a PBT is aligned with human judgment of property assertions, achieving a precision of 100\% and recall of 97\%. Finally, we evaluate the property coverage of LLMs across all API methods and find that the best model (GPT-4) is able to automatically synthesize correct PBTs for 21\% of properties extractable from API documentation.
\end{abstract}

\maketitle
\section{Introduction}

Property-based testing (PBT) is a powerful testing technique for testing properties of a program through generation of inputs. Unlike traditional testing methods that rely on manually written test cases and examples, PBT uses automatic generation of a wide range of inputs to invoke a diverse set of program behaviors. PBT was first popularized by the Quickcheck~\cite{claessen2000quickcheck} library in Haskell, and has been used to find a plethora of bugs in a variety of real-world software~\cite{arts2006testing, arts2015testing, hughes2016experiences, hughes2016mysteries}. Additional techniques %have been 
built on top of PBT~\cite{padhye2019jqf, maciver2019hypothesis, Lampropoulos19} %and 
have demonstrated %their 
potential in providing stronger testing for software.     

Despite its proven results and impact in the research community, PBT is not as widely adopted by open source and industry software developers. Using the Open Source Insights \cite{DepsDev} dependency dataset, we find that only 222 out of ~180,000 PyPI packages list the Python PBT library \emph{Hypothesis} as a dependency, despite it being a very popular project (6.7k+ stars on GitHub). Harrison et al.~\cite{goldstein2024property} conducted a series of interviews and detail a set of challenges faced by professional developers when attempting to use PBT in their software. Developers reported difficulties in (1) writing random data generators for inputs and (2) articulating and implementing properties that would meaningfully test their code. Furthermore, Harrison et al. describe the ``critical mass problem'' that PBT is still relatively unknown and unpopular among the software industry.

% \cl{possible writing pitch. Mention the existence of API documentation first, and how this could be extracted into PBT. Then, mention that LLMs are a possible way to extract API documentation into PBTs. See the first two paragraphs on the second column of the introduction of CodaMOSA for a flow there: \url{https://www.carolemieux.com/codamosa_icse23.pdf}}
% \cl{If following my previous suggestion, a nice transition between the last paragraph and this one is along the lines of : ``While developers have been reticent to adopt PBT, the practice of \emph{documenting code} is widespread.''}

\begin{figure}[t]
    \centering
    \includegraphics[width=0.9\columnwidth]{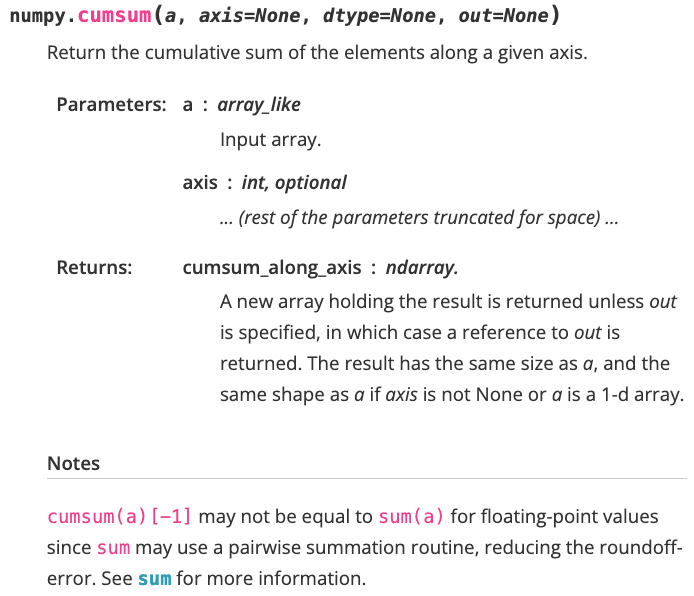}
    
    \caption{Truncated Numpy documentation for the \texttt{numpy.cumsum} API method. The documentation includes descriptions of properties about the result shape/size and additional information about the last element of the result.}
    \label{fig:numpy-doc}
\end{figure}

While developers have been reticent to adopt PBT, the practice of \emph{documenting code} is widespread. Documentation for library API methods is fairly common, especially  for  languages like Python, and contains valuable information about input parameters and properties of the output. An truncated version of the documentation for the \texttt{numpy.cumsum} API method can be seen in Figure~\ref{fig:numpy-doc}.  

% For example, Figure~\ref{fig:numpy-doc} displays a truncated version of the documentation for the \texttt{numpy.cumsum} method, which contains descriptions for each parameter (e.g. input array \texttt{a}) and properties about the resulting \texttt{cumsum(a)}. We particularly note the description that the "result has the same size as \textit{a}, and the same shape as \textit{a} if \textit{axis} is not None or \textit{a} is a 1-d array." This is natural language for a property that can be used in PBT. 

Recently, the use of pre-trained large language models (LLMs) for code generation has become increasingly popular~\cite{chen2021evaluating, fried2022synthesis, bubeck2023sparks}.  LLMs have been effective at translating natural language specifications and instructions to concrete code~\cite{ouyang2022training, openai2023gpt4}. Additionally, LLMs have shown potential to improve existing automated unit test generation techniques~\cite{lemieux2023codamosa}, and even generate unit tests from scratch~\cite{lahiri2022interactive,schafer2023testpilot}. %by providing example test cases to escaping coverage plateaus. 
In this paper, we investigate the potential of using LLMs to generate \textit{property-based tests} when provided API documentation. We believe that the documentation of an API method can assist the LLM in producing logic to generate random inputs for that method and deriving meaningful properties of the result to check. 

While LLMs have shown effectiveness in synthesizing unit tests~\cite{schafer2023empirical} and fuzz harnesses~\cite{zhang2023understanding, huang2024large, ossfuzz23}, synthesizing property-based tests has unique challenges. In unit testing, the test oracles are usually simple equality assertions between expected and actual outputs; in PBT, a property assertion that holds \textit{generally} for inputs must be synthesized in the test. Similarly, fuzz harnesses usually only contain \textit{implicit} oracles that capture unexpected behaviors such as uncaught exceptions or crashes. Further, synthesizing property-based tests requires generating custom logic for random input generation over various types of inputs; unit tests do not require this logic since there is only one input, and fuzz harnesses primarily rely on inputs represented as raw byte streams.

An example of using LLMs for PBT can be seen in Figure~\ref{fig:gpt-4-numpy-pbt}, which displays an LLM-synthesized property-based test for the \texttt{numpy.cumsum} method when provided the documentation in Figure~\ref{fig:numpy-doc}. First, the logic for generating random values for the input parameters \textit{a} and \textit{axis} is in lines~\ref{generator_begin}--\ref{generator_end}. Then, the \texttt{cumsum} method is invoked on these arguments on line~\ref{api_call}. Finally, lines~\ref{property_shape_begin}--\ref{property_result_end} contain property assertions for the output \texttt{cumsum\_result}. We specifically note that these properties assertions match natural language descriptions in the API documentation in Figure~\ref{fig:numpy-doc}. The documentation specifies that ``result has the same size as \textit{a}'', which has a direct translation to the assertion in line~\ref{property_size}. Similarly, the specification that result has ``the same shape as \textit{a} if \textit{axis} is not None or \textit{a} is a 1-d array'' is checked conditionally as an assertion in lines~\ref{property_shape_begin}--\ref{property_shape_end}. Finally, the property assertion shown in lines~\ref{property_result_begin}--\ref{property_result_end} checks that the last element of the result is equal to \texttt{np.sum(a)} if the array is not of \texttt{float} type. This assertion translates information from the notes section in the documentation into a useful property to check. While not a perfect property-based test, this example demonstrates the ability of LLMs to write logic for generating random inputs and derive meaningful property assertions from API documentation. 

\begin{figure}[t]
    \lstset{
    language=Python,
    basicstyle=\footnotesize\ttfamily,
    numbers=left,
    numberstyle=\tiny,
    numbersep=5pt,
    showstringspaces=false,
    framesep=2mm,
    xleftmargin=1.8em,
    tabsize=4,
    breaklines=true,
    breakatwhitespace=true,
    keywordstyle=\color{pythonblue},
    commentstyle=\color{pythongreen},
    stringstyle=\color{pythonpurple},
    identifierstyle=\color{black},
    numberstyle=\color{pythongray},
    escapeinside={!}{!},
}

\begin{lstlisting}
from hypothesis import given, strategies as st
import numpy as np

# Summary: Generate random input parameters for 
# numpy.cumsum and test properties
@given(st.data())
def test_numpy_cumsum(data):
    # Generating a list with varying length 
    # and integer elements
    a = data.draw(st.lists( !\label{generator_begin}!
        st.integers(min_value=-10, max_value=10),
        min_size=0, max_size=10))

    # Generate random axis
    axis = data.draw(st.one_of(st.none(),
        st.integers(min_value=0, 
                max_value=a.ndim-1)))  !\label{generator_end}!

    # Call numpy.cumsum with generated input
    cumsum_result = np.cumsum(a, axis=axis) !\label{api_call}!

    # Test property: output shape
    # should be the same as input shape 
    # if axis is not None or input is 1-d array
    if axis is not None or a.ndim == 1: !\label{property_shape_begin}!
        assert cumsum_result.shape == a.shape  !\label{property_shape_end}!

    # Test property: output size should be the 
    # same as input size
    assert cumsum_result.size == a.size !\label{property_size}!

    # Test property: cumsum(a)[-1] should be 
    # approximately equal to sum(a) for 
    # non floating-point values
    if not np.issubdtype(a.dtype, np.floating): !\label{property_result_begin}!
        np.testing.assert_almost_equal(
                cumsum_result.flatten()[-1], np.sum(a)) !\label{property_result_end}!
# End program
\end{lstlisting}
    \caption{A GPT-4 generated property based test for \texttt{numpy.cumsum}. The test first generates random integer arrays between size 1 and 20 and a random axis. Then, the API method under test \texttt{np.cumsum} is invoked on the randomly generated inputs. Finally, three properties are checked on the output array, all derived from information in the API documentation. All comments are also generated by GPT-4. }
    \label{fig:gpt-4-numpy-pbt}
\end{figure} 

In this paper, we study the use of state-of-the-art LLMs---Open AI's GPT-4, Anthropic's Claude-3-Opus, and Google's Gemini-1.5.-Pro---to automatically synthesize property-based tests. We propose single stage and two stage approaches for using API documentation to prompt the LLM to synthesize property-based tests. But how do we know whether an LLM-synthesized property-based test is good enough? We characterize several desirable properties of these tests and propose a methodology to evaluate the LLM's output: property-based tests must be \emph{valid} (free of compile-time or run-time errors), \emph{sound} (only assert properties that must be true), and ideally \emph{complete} (i.e., actually check for properties that are mentioned in the API documentation). To measure completeness of property assertions, we introduce the notion of \textit{property coverage}, which uses \emph{mutation testing} to measure the ability of a property-based test to fail when the target API method behaves in a way that violates its documented property. We evaluate our three models across two prompting strategies on 40 API methods across 10 Python libraries. We find that with the best-performing approach (GPT-4 with two-stage prompting), a valid and sound property-based test can be synthesized over 2.4 samples on average, and that a correct PBT can be synthesized for over 20\% of the documented properties. We thus find LLM-based PBT generation a suitable approach for automating some of the tedious process of writing property-based tests.

In summary, our contributions are the following:
% \cl{My higher-level take, re: my general dislike of being ``the first'': we make contribution 2 be the first contribution, and split the second into two contributions: (2) We propose three querying strategies to use LLMs, based on preliminary investigation (3) We outline a large-scale evaluation to evaluate both the suitability of LLMs for PBT generation, as well as the quality of our evaluation methodology }
\begin{enumerate}
    \item We propose an approach for using LLMs to synthesize property-based tests given API documentation.
    \item We propose a methodology for evaluating LLM-synthesized property-based tests considering validity, soundness, and completeness.
    \item We demonstrate alignment between our metric for soundness and human judgment through manual labeling. 
    \item We propose a novel metric of \emph{property coverage} for measuring completeness with respect to documented properties.
    \item We present an empirical evaluation of the validity, soundness, and property coverage of PBTs synthesized by three state-of-the-art commercial LLMs across two prompting strategies.
\end{enumerate}

\section{Background}
\subsection{Property-based Testing}
Property-based testing~\cite{claessen2000quickcheck} aims to probabilistically test a program by generating a large number of random inputs and checking whether the corresponding outputs of the program adhere to a set of desired properties. A property-based test can be defined as the following: given a function/method under test $f$, an input space $\Chi$, and a property $P$, we want to validate that $ \forall x \in \Chi : P(x, f(x)) $. Often, $P$  is a conjunction %comprises a collection
of component properties, that is, $P = p_1 \land p_2 \land \dots p_k$.
In practice, we are unable to enumerate all inputs in $\Chi$. So, we write a \emph{generator} function \textit{gen} that %samples from $\Chi$ to 
produces a random input in $\Chi$, i.e. $x = \textit{gen}()$. Then we write a \textit{parametrized test} $T :: X \rightarrow \{\textit{true}, \textit{false}\}$ that returns $P(x, f(x))$.  The property is checked on many randomly generated %values 
$x$. %and a
A violation of the property causes the test to fail. Although PBT cannot prove the absence of property violations, it improves %is %nevertheless an improvement 
over testing specific hard-coded inputs and outputs as is commonly done in unit testing. PBT is thus a form of \emph{fuzz testing}.

% \paragraph{Generators}. The developer must write a function to generate random data inputs to the program. Writing a generator is a large barrier of entry to property-based testing, as it requires reasoning about the entire input space to the program compared to individual input/output pairs in a unit test. Additionally, the type of input can range in complexity from simple literals to entire programs. 
% \paragraph{Properties}. The developer must identify and writing assertions for the desired properties on the output of the program.  

% As an example, suppose we would like to write a property-based test for the Python \texttt{sorted} function on lists.  

% \begin{figure}[t]
%     \centering
%     \includegraphics[height=4.5cm]{figures/pbt_loop.pdf
%     }
%     \caption{A property-based testing loop. A generator is used to produce random input data that are executed in the program under test. Properties of the expected behavior of the program are checked for each randomly generated input. Any assertion failure that occur due to the violation of the property checks are reported to the user. }
%     \label{fig:pbt}
% \end{figure}

%\subsubsection{Hypothesis}
\begin{figure}[t]
\lstset{
    language=Python,
    basicstyle=\footnotesize\ttfamily,
    numbers=left,
    numberstyle=\tiny,
    numbersep=5pt,
    showstringspaces=false,
    framesep=2mm,
    xleftmargin=1.8em,
    tabsize=4,
    breaklines=true,
    breakatwhitespace=true,
    keywordstyle=\color{pythonblue},
    commentstyle=\color{pythongreen},
    stringstyle=\color{pythonpurple},
    identifierstyle=\color{black},
    numberstyle=\color{pythongray},
    escapeinside={!}{!},
}

\begin{lstlisting}
from hypothesis import given, strategies as st

# Random list generator
@st.composite !\label{sorted_st_composite}!
def generate_lists(draw):
  return draw(st.lists(elements=st.integers(), !\label{generator_lists_separate}!
                      min_size=1)) !\label{min_size_1}!

# Property-based test for sorted with 
# separate generator
@given(lst=generate_lists()) !\label{sorted_given_separate}!
def test_sorted_separate(lst): 
  sorted_lst = sorted(lst) !\label{sorted_invoke_separate}!
  assert all(sorted_lst[i] <= sorted_lst[i + 1] !\label{sorted_assertion_separate_begin}!
              for i in range(len(sorted_lst) - 1)) !\label{sorted_assertion_separate_end}!

# Alternative property-based test 
# with inline generator
@given(st.data()) !\label{sorted_st_data}!
def test_sorted_combined(data):
  lst = data.draw(st.lists(elements=st.integers(), !\label{generator_lists_combined}!
                             min_size=1)) !\label{min_size_2}!
  sorted_lst = sorted(lst) !\label{sorted_invoke_combined}!
  assert all(sorted_lst[i] <= sorted_lst[i + 1] !\label{sorted_assertion_combined_begin}!
                  for i in range(len(sorted_lst) - 1)) !\label{sorted_assertion_combined_end}!
    \end{lstlisting}
    \caption{Example property-based tests in Hypothesis for the Python \texttt{sorted} function to sort lists. The \texttt{test\_sorted\_separate} function uses a separate generator, whereas the function \texttt{test\_sorted\_combined} combines 
    the generator and testing logic into one function.}
    \label{fig:hypothesis-lists}
\end{figure}

We next describe how our formal definition %of a %property-based test 
translates to %property-based testing 
code in Hypothesis~\cite{maciver2019hypothesis}, a popular PBT library for Python. 

Suppose our function under test is the Python \texttt{sorted} function, which takes in a \texttt{list} as input and returns a sorted version. We want %would like 
to test the property that the elements of the sorted list are monotonically increasing. 

First, we must write our generator \textit{gen} that samples an input from the input space of lists. An example of such a generator is the \texttt{generate\_lists} function in lines~\ref{sorted_st_composite}--\ref{min_size_1} of Figure~\ref{fig:hypothesis-lists}. Hypothesis has a built-in set of sampling \textit{strategies} for various data structures. The \texttt{lists} and \texttt{integers} strategies in line~\ref{generator_lists_separate} are used to randomly generate and return a Python integer list of size $\geq$ 1.

Next, we must write the parametrized test $T$ that takes in an input $x$ and returns $P(x, f(x))$, where $f$ is the \texttt{sorted} function 
and $P$ is the property that the elements of the sorted list are monotonically increasing. An example of such a parametrized test is %the 
\texttt{test\_sorted\_separate} %function seen 
in Figure~\ref{fig:hypothesis-lists}. In line 12, the \texttt{sorted} function is invoked on the input \texttt{lst}. Then, lines~\ref{sorted_assertion_separate_begin}--\ref{sorted_assertion_separate_end} check the property $P$ that elements of the sorted listed are increasing by using an \textit{assertion} statement. $T$ %will 
returns \textit{true} if $P(x, f(x))$ is true, i.e. there is no assertion failure. Generally, if $P$ %were to consist of 
had multiple component properties, %it can be represented as 
there would be
a list of assertion statements in $T$.

Finally, to complete the property-based test, we must invoke our generator to sample random inputs and call the parametrized test on the input. This is done using the Hypothesis \texttt{@given} decorator, as seen in line~\ref{sorted_given_separate}. The decorator specifies that the input \texttt{lst} of our parametrized test \texttt{test\_sorted\_separate} should use \texttt{generate\_lists} as the generator. 

Another style of writing a Hypothesis test is to include the generator inside the parametrized test, as seen in the function \texttt{test\_sorted\_combined} in Figure~\ref{fig:hypothesis-lists}. At line~\ref{sorted_st_data}, the \texttt{@given(data())} decorator provides an object which can be used to sample random input data of unspecified type. Lines~\ref{generator_lists_combined}--\ref{min_size_2} act as the generator, using the same logic as the \texttt{generate\_lists} function to a generate random integer list of with minimum size 1. Lines~\ref{sorted_invoke_combined}--\ref{sorted_assertion_combined_end} use the method invocation and assertion statements as is in \texttt{test\_sorted\_separate}. The approach of including the generator in the parametrized test has particular advantages when the method under test has multiple input parameters that have dependencies with each other. In this scenario, each argument can be sequentially generated one at a time using generators that depend on previously generated arguments.

While the property-based tests shown in Figures~\ref{fig:hypothesis-lists} are valid and will properly run, they are not necessarily the \textit{best} property-based tests for the \texttt{sorted} function. Perhaps the user would like validate the behavior of \texttt{sorted} on the empty list, which is not an input produced by our generator due to the \texttt{min\_size=1} constraint in lines~\ref{min_size_1} and \ref{min_size_2}. Similarly, the assertions in lines \ref{sorted_assertion_separate_begin}--\ref{sorted_assertion_separate_end} and lines~\ref{sorted_assertion_combined_begin}--\ref{sorted_assertion_combined_end} do not capture all behavior of the \texttt{sorted} function. For instance, it does not check that \texttt{lst} and \texttt{sortedlst} share the same elements. We discuss these types of challenges more in Section~\ref{sec:proptest-ai:eval}.   

\subsection{Large Language Models}
Pre-trained large language models (LLMs)~\cite{shoeybi2019megatron, brown2020language, chowdhery2022palm, thoppilan2022lamda, openai2023gpt4, reid2024gemini, claude3} are a class of neural networks with a huge number of parameters, trained on large corpora of text data.
These models are %typically 
trained in an \emph{autoregressive} manner---i.e., trained to predict the next token in a sequence---which allows them to be trained on a large volumes of unlabelled text. This extensive pre-training allows them to function as \emph{one-shot} or \emph{zero-shot} learners~\cite{brown2020language}. That is, these models can perform a variety of tasks when given only one example of the task, or a textual instruction of the tasks. The natural-language instructions, along with any additional input data, that are passed to the LLM are called the \emph{prompt}~\cite{liu2023pre}.
The practice of creating prompts that allow the LLMs to effectively solve a target task is called \emph{prompt engineering}.

Further, a number of LLMs have been trained extensively on code~\cite{chen2021evaluating,xu2022polycoder,li2023starcoder,roziere2023code,guo2024deepseek}. These models, as well as more general-purpose LLMs, have been used for numerous software engineering tasks, including program synthesis~\cite{austin2021synthesis,fried2022synthesis}, program repair~\cite{prenner2021progrepair,pearce2021fixbugs,pearce2023vulrepair}, code explanation~\cite{sarsa2022explanations}, and test generation~\cite{lemieux2023codamosa,lahiri2022interactive,schafer2023testpilot}. These techniques use the LLMs out-of-the-box, getting them to accomplish the tasks via prompt engineering alone. 

Like prior work, we use pre-trained language models and adapt them to our tasks only via prompt engineering. We discuss our methods of constructing these prompts in Section~\ref{sec:proptest-ai}.

%  The benefit of using prompts for LLMs is that it does not require any retraining of the model, which is a very expensive process. Through careful construction of prompts (\textit{prompt engineering}), LLMs can be applied to diverse coding tasks such as code completion and code repair. 
%  In our approach, we employ prompt engineering to generate property-based tests from the GPT-3.5 and GPT-4 models.
% \todo{add more about GPT-3.5 and GPT-4}

\section{The Proptest-AI Approach} 
\label{sec:proptest-ai}
\subsection{Prompt Design}
\label{sec:proptest-ai:prompt-design}

To synthesize a property-based test from the LLM, we first design prompts that include the API documentation and instructions to write a property-based test for the input method. We explored two prompting strategies: one to generate a single property-based test and one to generate a property-based testing suite.
 
%We begin by designing a high-level prompt template composed of the following parts:
Our high-level prompt templates contains:
\begin{enumerate}
    \item System-level instructions stating that the LLM is an expert Python programmer.
    \item The target
    API documentation, taken from the API website.
    \item User-level task instructions to review the API documentation and perform a particular task.
    \item The desired output format.
\end{enumerate}

\begin{figure}[t]
\centering
\includegraphics[width=0.8\columnwidth]{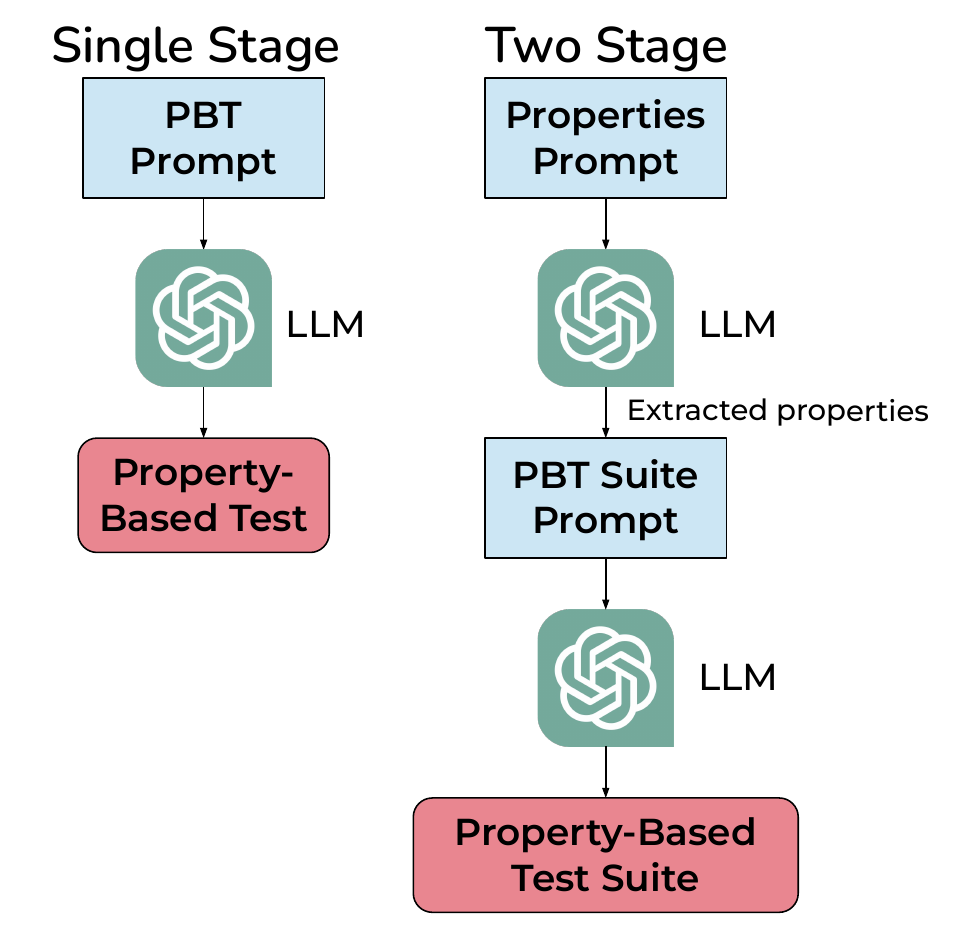}
\caption{Two methods of generating property-based test using an LLM. The first is a single stage prompt of the LLM with zero-shot CoT instructions to (1) explain a generation strategy, (2) properties to test, and (3) generate a single property-based test. The second method instructs the LLM to extract a list of properties from the API docs and \textit{continues the conversation}, instructing the LLM  to write a test for each property.}
\label{fig:proptest-ai-approaches}
\end{figure}

Based on this template, our first prompting strategy generates a single property-based test. The user-level task instructions begin with Chain of Thought~\cite{kojima2022large} instructions to outline a generation strategy and a list of properties to test before synthesizing the property-based test. The prompt ends with instructions to generate a Hypothesis PBT and adds an output format including Hypothesis import statements, relevant API import statements, a boilerplate function signature, and an example Hypothesis PBT from the Hypothesis documentation. We refer to this prompt to generate the entire property-based test as the ``PBT Prompt'', as seen in Figure~\ref{fig:proptest-ai-approaches}. 

% An example of a prompt following this design to generate a single property-based test for the \texttt{networkx.find\_cycle} method can be seen in Figure~\ref{fig:proptest-ai-prompt}. The user-level tasks instruct to review the API documentation for \texttt{find\_cycle} and write a function that generates random values of the \texttt{networkx.Graph} object. The output format uses the \texttt{st.data} decorator (ref.~\ref{sorted_st_data} of Figure~\ref{fig:hypothesis-lists}).

The second prompting strategy we explore has a two stage prompting method of (1) instructing the LLM to extract properties from the API (maximum five) and (2) instructing the LLM to create individual property-based test functions for each property. The first prompt contains user-level task instructions to extract five properties that hold for all outputs of the API method. The second prompt contains instructions to synthesize a PBT using Hypothesis for each of the properties generated by the LLM. Our full prompt templates are available in our data artifact.

% \begin{figure}[t]
% \centering
% \includegraphics[width=0.8\linewidth]{figures/example_prompt_text.pdf}
% \caption{An example prompt template for synthesizing a single property-based test for the \texttt{networkx.find\_cycle} API method.}  
% \label{fig:proptest-ai-prompt}
% \end{figure}

\subsection{Evaluation Methodology}
\label{sec:proptest-ai:eval}
Using our Proptest-AI methodology to prompt the LLM to synthesize property-based tests, how do we evaluate the quality of these generated PBTs? While the effectiveness of unit tests has been a well studied topic for decades~\cite{goodenough1975toward, frankl1998further, fraser2014large, shamshiri2015automatically}, this is not the case for property-based tests. One difficulty in conducting these types of evaluations for PBT is the lack of readily available property-based tests for software. Thankfully, LLMs can provide us a method of automatically generating property-based tests for which we can design an evaluation methodology.

% Metrics such as code coverage and mutation score~\cite{lukasczyk2022pynguin, lemieux2023codamosa} have been used as a proxy for automated unit test generation effectiveness. To evaluate unit test oracles, previous approaches have measured similarity to a ground-truth human written oracle \cite{watson2020learning, tufano2022generating, dinella2022toga}. 

We propose a PBT evaluation methodology and metrics focusing on (1) the validity of the tests, (2) the soundness of the tests, and (3) the \textit{property coverage} (detailed in Section~\ref{sec:prop-coverage}) of the tests. To motivate each of these metrics, we include examples of inaccurate LLM-synthesized PBTs %for API methods 
and discuss the issues that impact each of these qualities. We then propose mitigation strategies for each of the inaccuracies that improve the quality of the property-based tests. All of our examples use the Hypothesis PBT library in Python. 
\subsubsection{Validity}
% First, we divide the evaluation of generator quality into \textit{generator validity} and \textit{generator diversity}.
\begin{figure}[t]
\begin{lstlisting}[style = python]
from hypothesis import strategies as st
from datetime import timedelta

@given(days=st.integers(min_value=0),
   seconds=st.integers(min_value=0),
   microseconds=st.integers(min_value=0),
   milliseconds=st.integers(min_value=0),
   minutes=st.integers(min_value=0),
   hours=st.integers(min_value=0),
   weeks=st.integers(min_value=0))
def test_timedelta_total_seconds(days, seconds, 
    microseconds, milliseconds, minutes, 
    hours, weeks):
  td = timedelta(days=days, seconds=seconds !\label{timedelta_constructor}!
    microseconds=microseconds, 
    milliseconds=milliseconds, 
    minutes=minutes,
    hours=hours, 
    weeks=weeks)
  (...)
    \end{lstlisting}
    \caption{An example invalid \texttt{datetime.timedelta} PBT produced by GPT-4. The \texttt{datetime.timedelta} constructor on line~\ref{timedelta_constructor} raises an \texttt{OverflowError} when the absolute magnitude of days exceeds 1,000,000.}
    \label{fig:gpt-generator-runtime-error}
\end{figure}

\label{sec:validity}
One type of incorrect behavior in a property-based test is a validity issue in which a run-time error is encountered when the test is executed. The problem of hallucination is a well-known problem with using LLMs to generate code~\cite{fan2023large}. LLMs may generate plausible-looking but incorrect code that throws unexpected errors at runtime. This issue arises in property-based testing as well, in which the LLM may synthesize test cases that are syntactically valid but result in a runtime error during execution. Since the test is executed on multiple random inputs, it is also possible that only a percentage of randomly generated inputs result in runtime errors. An example is the generator in Figure~\ref{fig:gpt-generator-runtime-error} for \texttt{timedelta} objects in the Python \texttt{datetime} module. The generator function can produce values for the \texttt{timedelta} object constructor that may result in an \texttt{OverflowError} raised by the \texttt{datetime.timedelta} constructor when the magnitude of days exceeds 1,000,000. Thus, we determine a property-based test as \textit{valid} if 100\% of test function invocations do not result in any run-time errors, excluding property assertion errors as these relate to soundness. 

\subsubsection{Soundness}
\label{sec:soundness}
LLMs may also synthesize a property-based test that is \textit{unsound}, i.e., there exists an input/output pair that violates a property assertion but is valid given the specification. Figure~\ref{fig:gpt-unsound-property-numpy-array} provides an example of an LLM-synthesized property-based test for the \texttt{numpy.cumsum} method that contains an unsound property on line~\ref{unsound_property_numpy}. The \texttt{numpy.cumsum} documentation specifies that for a given input array $a$, the output should have "the same shape as $a$ if axis is not None or $a$ is a 1-d array". The synthesized property is unsound because it unconditionally checks whether the output and input shapes match. A randomly generated input of \texttt{array([[0]]} produces an assertion failure when this test is run since the input shape is \texttt{(1, 1)} and the output shape is \texttt{(1,)}; this is not an actual bug and the behavior of the \code{cumsum} method conforms with the API documentation.

If we encounter an assertion failure from a property check during the test, how do we know whether it is due to an unsound property or due to a bug in the API implementation? Given an assertion failure, we assume that the likelihood of the LLM generating an unsound property is higher than the likelihood of a bug. This is a weak assumption. We can capture the soundness of a property by measuring the frequency at which the assertion fails across multiple generated inputs. If an assertion fails on a significant percentage of the inputs, it is most likely an unsound property. 

A property-based test is determined as \textit{sound} if 100\% of test function invocations do not result in assertion errors from the property checks. We specifically filter out any executions that result in any other runtime errors, as these are related to the validity of the property-based test rather than the soundness.  

% \paragraph{Refinement Strategy:} Similar to fixing invalid property-based tests, we construct a prompt to iterate upon an unsound property-based test. When encountering an assertion error, Hypothesis will report the specific input/output example that resulted in the failure. Our prompt includes the input to the API method that caused the assertion failure, the expected and actual outputs of the arguments in the property assertion, and instructions to fix the property assertion. This prompt template is a slight modification to Figure~\ref{fig:fix-prompt} that additionally includes the input/output example.

% We used ran ten GPT-synthesized property-based test 10,000 times for the \texttt{numpy.cumsum} and found that 68\% of the properties were sound. 6 of the 10 synthesized property-based tests did not contain any unsound property assertions. The majority of the unsound properties contained an unconditional equality check between \texttt{cumsum(a)[-1]} and \texttt{np.sum(a)}, which is not necessarily true for floating point values (ref. Figure~\ref{fig:numpy-doc}).

% A mitigation strategy for unsound properties is to feed back the error message along with an input that results in the assertion error to the LLM. After running the property-based test, Hypothesis reports a falsifying example input that causes the assertion to fail. \cl{I don't understand the point of the last sentence here}

\begin{figure}[t]
\begin{lstlisting}[style = python]
@given(generate_array())
def test_cumsum(a):
  # Test that the shape of the output 
  # is the same as the input
  out_shape = np.cumsum(a).shape
  assert out_shape == a.shape !\label{unsound_property_numpy}! 
    \end{lstlisting}
    \caption{A Gemini-1.5-Pro generated property-based test containing an unsound property for \texttt{numpy.cumsum} on line~\ref{unsound_property_numpy}.}
    \label{fig:gpt-unsound-property-numpy-array}
\end{figure}
\subsubsection{Property Coverage}

% \begin{figure}[t]
% \centering
% \includegraphics[width=0.8\linewidth]{figures/fix_prompt_template.pdf}
% \caption{Prompt template for fixing invalid property-based test for the \texttt{networkx.find\_cycle} API method.}  
% \label{fig:fix-prompt}
% \end{figure}

\label{sec:prop-coverage}
% Validity and soundness provide a measure of whether the property-based test is able to execute without errors, which is essential for a good property-based test. However, a property-based test should also effectively test the properties  may be sound, it may be a \textit{weak} assertion, i.e. it does not properly test a property. However, we need an additional measurement capturing the effectiveness of the PBT at testing properties. We formulate the Specifically: if the API method were return an output that \textit{violated} a given property, would the property-based test detect this violation?   

\begin{figure}[t]
    \lstset{
    language=Python,
    basicstyle=\footnotesize\ttfamily,
    numbers=left,
    numberstyle=\tiny,
    numbersep=5pt,
    showstringspaces=false,
    framesep=2mm,
    xleftmargin=1.8em,
    tabsize=4,
    breaklines=true,
    breakatwhitespace=true,
    keywordstyle=\color{pythonblue},
    commentstyle=\color{pythongreen},
    stringstyle=\color{pythonpurple},
    identifierstyle=\color{black},
    numberstyle=\color{pythongray},
    escapeinside={!}{!},
}

\begin{lstlisting}
import numpy as np

# Violate Non-Decreasing Sequence
def buggy_cumsum_1(a, 
    axis=None, dtype=None, out=None):
  result = np.cumsum(a, axis=axis, dtype=dtype, 
                        out=out)
  # Reverse the result to create decreasing elements
  return result[::-1]
    \end{lstlisting}
    \caption{Example property mutant of \texttt{numpy.cumsum} produced by GPT-4. The mutant violates the property that the output of \texttt{numpy.cumsum} must be non-decreasing by reversing the result.}
    \label{fig:property-mutant}

\end{figure}

While validity and soundness indicate that a property-based test runs without errors, an additional measurement is needed to evaluate effectiveness at testing specific properties. A property-based test may execute correctly with 100\% validity and 100\% soundness, yet inadequately test key properties due to having weak assertions. Our goal is to establish a metric that answers the question: how effective is the property-based test at detecting whether an API method violates a specific property?

One related idea is \textit{mutation testing}, which measures the ability of a test to detect bugs artificially bugs in the program under test~\cite{DeMillo78}. First, mutant versions of the program are created with an artificially injected bug. If the test fails when executing the mutated programs, then the mutant is \textit{killed}; otherwise, the mutant \textit{survives}. Mutation testing provides a measurement related to the bug-finding capability of a particular test.

Mutation testing injects bugs by applying syntactic operators on the \textit{source code}; these types of syntactic operators may not necessarily correspond to property violations of the API method. We would like to construct a mutant that performs a property-level mutation on the API method. More formally, given a method under test $f$, we would like to generate a mutant $f^\prime$ such that $\forall x \in \Chi : \neg P(x, f^\prime(x)) $. 

Thus, we define a \textit{property mutant} as a buggy version of the API method that returns an output violating a specific property. A property mutant $f^\prime$ of a method under test $f$ is defined as follows: 
$$ f^\prime(x) = \textit{mut}(f(x))$$
where \textit{mut} is a mutation operation on $f(x)$. The property mutant contains the same signature as the API method, invokes the original API method on the input, and finally performs a mutation operation on the output to violate the property. Generating this mutation operation requires semantic reasoning about how an output would violate a property.

%Recently, 
LLMs have been used to perform %shown promise in performing
bug injection and create program mutants~\cite{ibrahimzada2023automated, gargcoupling}. We design a specific prompt for the LLM to generate property mutants of a method under test for a given property. An example of a property mutant for \texttt{numpy.cumsum} is shown in Figure~\ref{fig:property-mutant}. This mutant can return an output that violates the property that the output must be non-decreasing by reversing the order of the output. In order to kill this mutant, the PBT must contain an assertion checking this property and contain logic to generate inputs of length greater than 1.

To check whether a property mutant is killed, we create a modified version of the property-based test. This modified version substitutes the call to the original API method with the call to the mutant API method. Assuming the original property-based test is valid and sound, we check whether there is an assertion failure on the output of the mutant API method in the modified property-based test. 

We define the \textit{property mutation score} of a property-based test as the percentage of killed property mutants of a given API method. The process of generating property mutants and measuring property mutation score is described as follows: 
\begin{enumerate}
    \item Prompt the LLM to extract a list of properties from the API documentation.
    \item Prompt the LLM to generate a set of property mutants for each property. 
    \item If the property-based test is sound, execute the property-based test with a substituted call to the property mutant API method instead of the original API method.  
    \item Check whether the property assertions fail in the modified property-based test. If so, then the mutant is killed.
\end{enumerate}
Finally, a property is considered \textit{covered} if the PBT is able to kill any of the corresponding property mutants. The overall \textit{property coverage} of a PBT is the percent of properties for which the PBT is able to kill property mutants.

\section{Evaluation}
\label{sec:eval}
Based on our proposed evaluation methodology, we structure our evaluation around three research questions:

\rqn{1}{Are LLMs able to synthesize valid and sound property-based tests of API methods when provided documentation?}

\rqn{2}{Is the execution-based soundness metric for property assertions aligned with human judgment of soundness?}

\rqn{3}{Are LLMs able to synthesize property-based tests that cover documented properties of API methods?}

\subsection{Experimental Setup}
\paragraph{Models}
We chose three state-of-the-art language models for our evaluation: OpenAI's GPT-4~\cite{achiam2023gpt}, Anthropic's Claude-3-Opus~\cite{claude3}, and Google's Gemini-1.5-Pro~\cite{reid2024gemini}. \footnote{%Although 
We initially included Codellama-34B in our models, but preliminary experiments showed that the validity of generated PBTs was too low to warramt further evaluation.}

\paragraph{API Methods}
We selected API methods from 10 different Python libraries to generate property-based tests across a variety of tasks and input data types. Table~\ref{tab:api-methods} displays the libraries and the number of selected API methods for each library. The libraries include native Python libraries as well as popular third-party libraries such has \texttt{pandas} and \texttt{numpy}. All selected API methods are deterministic, as this is necessary for property-based testing. The API method documentation was extracted from the online documentation for each API method. In total, we selected 40 API methods. The full list of API methods and documentation is provided in the data artifact.

\begin{table}[t]
\caption{Modules selected for Proptest-AI evaluation. Selected modules consist of native Python libraries (e.g. \texttt{datetime} and \texttt{statistics}) and third-party libraries (e.g. \texttt{networkx} and \texttt{numpy}). The number of API methods selected from each library as well as average token length of the API method documentation, using the OpenAI tokenizer.}
\small
\begin{tabular}{lll}
\toprule
\multirow{2}{*}{\textbf{Library}} & \textbf{\# API} & \multirow{1}{*}{\textbf{Documentation}} \\
                 &     \textbf{Methods}                   & \multirow{1}{*}{\textbf{Length (tokens)}} \\
\midrule
\texttt{dateutil} & 2 & $643.5\pm 155.5$ \\
\texttt{html} & 2 & $81.5\pm 5.5$ \\
\texttt{zlib} & 3 & $312.7\pm 116.8$ \\
\texttt{cryptography.fernet} & 3 & $458.7\pm 100.9$ \\
\texttt{datetime} & 4 & $212.8\pm 83.2$ \\
\texttt{decimal} & 5 & $122.2\pm 68.8$ \\
\texttt{networkx} & 5 & $592.8\pm 439.6$ \\
\texttt{numpy} & 5 & $766.2\pm 242.4$ \\
\texttt{pandas} & 5 & $1197.8\pm  569.7$ \\
\texttt{statistics} & 6 & $318.6\pm 149.5$ \\
\midrule
\textbf{Total} & 40 & - \\
\bottomrule
\end{tabular}
\label{tab:api-methods}
\end{table}

\paragraph{Approaches}
We evaluate two methods of prompting the LLM to generate a property-based test shown in Figure~\ref{fig:proptest-ai-approaches}. (1) A single-stage method to generate a single test function and (2) a two-stage method to extract properties and generate a property-based test suite of test functions. With three models and two prompting methods, we have a total of 6 approaches. 

\paragraph{Samples}
For each approach, we sample the LLM five times with a temperature of 0.7. In total, we synthesized 1,200 samples across all API methods and approaches. Out of these samples, only 16 either contained no Python or were syntactically invalid Python code.

\subsection{RQ1: Validity and Soundness}
\label{sec:eval:validity}

\begin{table*}[t]

\small
\centering
\caption{Proptest-AI synthesized valid and sound test functions across all API methods. 41.74\% of the test functions synthesized using the two-stage prompting approach with GPT-4 achieve 100\% validity and 100\% soundness.}
\begin{tabular}{llcccc}
\toprule
\textbf{Model} & \textbf{Approach} & \textbf{Total Test Functions} &\textbf{Valid Test Functions} & \textbf{Valid and Sound Test Functions} \\ \midrule
\multirow{2}{*}{Claude-3-Opus} & Single Stage & 215 & 53 (24.65\%) & 16 (7.44\%) \\
       & Two Stage & 931 & 423 (45.43\%)& 289 (31.04\%) \\ \midrule
\multirow{2}{*}{Gemini-1.5-Pro} & Single Stage & 220 & 39 (17.72\%) & 25 (11.36\%) \\
       & Two Stage & 799 & 209 (26.16\%) & 124 (15.51\%) \\ \midrule
\multirow{2}{*}{GPT-4} & Single Stage & 212 & 97 (45.75\%) & 54 (25.47\%) \\
       & Two Stage & 733 & \textbf{400 (54.57\%)} & \textbf{306 (41.74\%)} \\ \midrule
\textbf{Total} & - & 3,110 & 1,221 (39.26\%) & 814 (26.17\%) \\ \bottomrule
\end{tabular}
\label{tab:valid-sound}
\end{table*}

% \begin{table*}
% \small
% \begin{center}
%     \begin{tabular}{llrr|rr}
%         \toprule
%          \multirow{2}{*}{\textbf{Model}} & \multirow{2}{*}{\textbf{Approach}}  & \multicolumn{2}{c|}{\textbf{valid@}} & \multicolumn{2}{c}{\textbf{sound@}} \\
          
%           & & 1 & 5 & 1 & 5 \\
%         \midrule
%         \multirow{2}{*}{Gemini-1.5-Pro}
%         & Single Stage & 19.15\% & 34.04\% & 40.43\% & 12.77\% \\
%         & Two Stage &  \textbf{29.79\%} & \textbf{42.55\%} & \textbf{51.06\%} & \textbf{17.02\%} \\
%         \midrule
%         \multirow{2}{*}{Claude-3-Opus}
%         & Single Stage & 32.00\% & 40.00\% & 44.00\% & \textbf{26.00\%} \\
%         & Two Stage & \textbf{34.00\%} & \textbf{48.00\%} & \textbf{50.00\%} & \textbf{26.00\%} \\
%         \midrule
%         \multirow{2}{*}{GPT-4}
%         & Single Stage & 23.40\% & 29.79\% & 29.79\% & 19.15\% \\
%         & Two Stage & \textbf{25.53\%} & \textbf{34.04\%} & \textbf{38.30\%} & \textbf{23.40\%} \\

%         \bottomrule
%         \end{tabular}
%         \caption{Average valid@k and sound@k aggregated across 40 Python API methods. valid@k refers to the percentage of test functions that achieve 100\% validity, and sound@k refers to the percentage of test functions that achieve 100\% soundness.}
%         \label{tab:valid-sound}
% \end{center}
% \end{table*}

High validity and soundness for property-based tests are important for ensuring that the test can run without errors (e.g., calling a nonexistent API or missing an import for a library). For RQ1, we report the validity and soundness of Proptest-AI synthesized PBTs across all API methods, models, and approaches. To measure validity, we call each property-based test function 1,000 times and check for non-assertion errors; for soundness, we check for assertion errors. 

Table~\ref{tab:valid-sound} shows the percent of valid and sound test functions across all API methods for the single-stage and two-stage approaches. We find that the two-stage prompting achieves significantly higher validity than the single-stage generation for all models. This is likely due to the test functions in the suite being smaller and only testing one specific property, thus having a smaller chance of hallucination. Similarly, the soundness of test functions synthesized using the two-stage prompting is much higher.

\begin{figure}[t]
\includegraphics[width=\columnwidth]{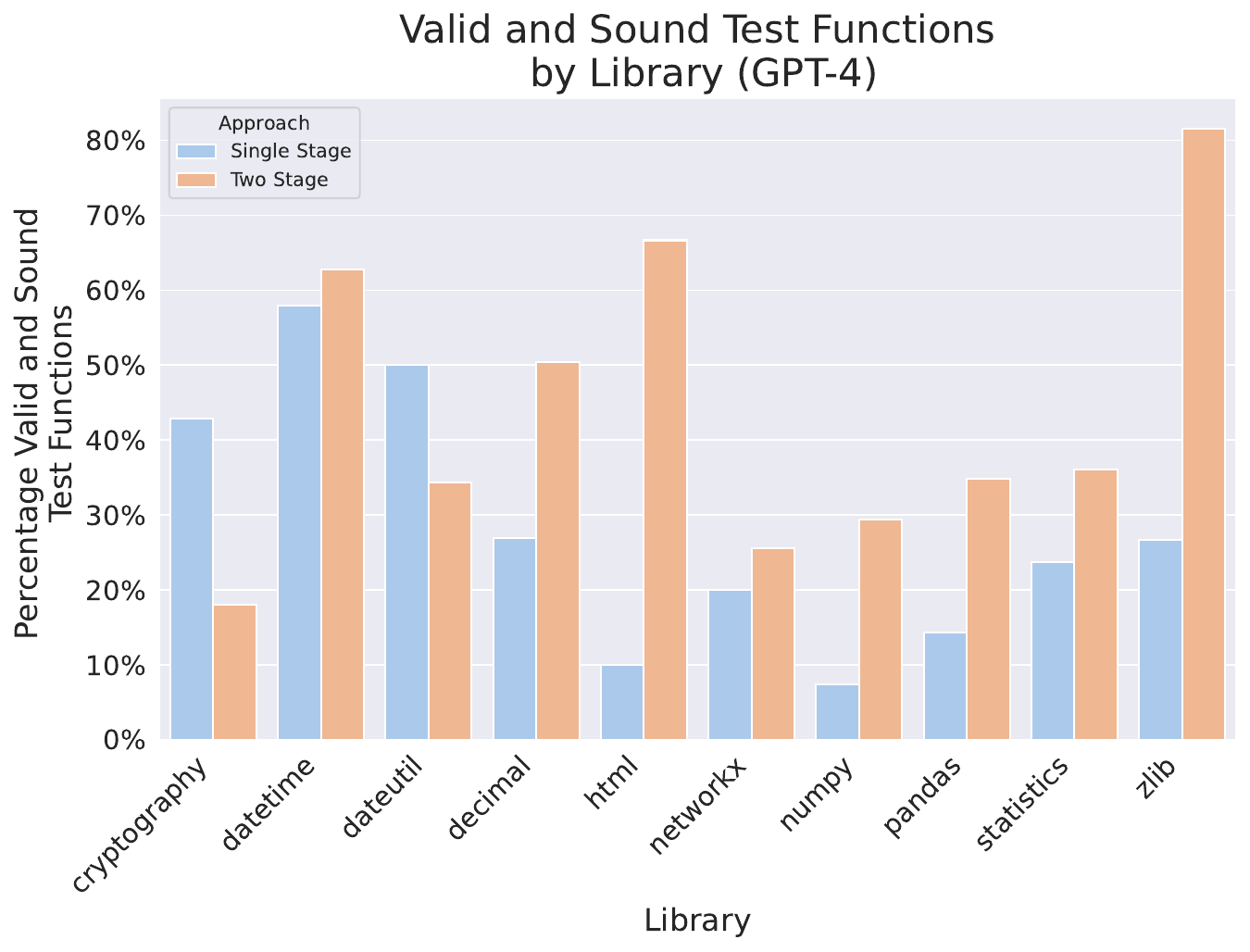}
\caption{Distribution of valid and sound property-based test functions for GPT-4 across all libraries. Certain libraries such as \texttt{networkx} and \texttt{pandas} have a smaller percentage of valid and sound test functions than \texttt{zlib}.}
\label{fig:valid-sound-dist}
\end{figure}

Overall, GPT-4 achieves the highest validity and soundness for both single-stage and two-stage prompting. The two-stage approach with GPT-4 is able to synthesized a valid and sound test function within 2.4 samples on average.  Figure~\ref{fig:valid-sound-dist} shows the distribution of valid and sound test functions from GPT-4 across our 10 different libraries. We observe the performance varying across libraries, with validity and soundness much higher on libraries such as \texttt{datetime} and \texttt{zlib}.  

We additionally report average soundness over each test function as the percentage of 1,000 invocations that do not result in any assertion errors. Figure~\ref{fig:soundness} shows the this distribution over all synthesized test functions. Although only 814 out of 1,221 valid test functions achieve 100\% soundness (Table~\ref{tab:valid-sound}), this distribution shows that the soundness of many of these tests functions close to 100\%. This suggests that the property assertions are correct for most inputs and outputs, but may fail to capture potential edge cases.

\mybox{The best Proptest-AI approach with two-stage prompting and GPT-4 is able to synthesize valid and sound property-based test functions in \textbf{2.4 samples} on average.}

% yes, sorry I'm removing the @k and need to re-edit these
%\cl{Can skip the "similar to validity..." if space is needed. I am guessing you are going to edit this sentence/add more text, otherwise it's weird to introduce the atK here.. :)}
%:+1:
% Similar to validity, we define \textit{soundness@k} as the maximum soundness achieved in the first $k$ samples of the LLM. \cl{Something is off about this sentence, but I am not sure enough about the intended meaning to edit myself}

\subsection{RQ2: Soundness Metric Alignment}
Our soundness metric reported in RQ1 provides an execution-based measurement for property assertions. However, it is possible that certain property assertions in the test are not executed due to certain program paths taken during execution. To further validate the soundness of individual property assertions, we performed a manual labeling of a sample of synthesized property-based test assertions. Out of all valid property-based test samples, half were selected, evenly distributed for each library. Raters read the documentation for the corresponding API method and labeled the first five assertions as sound or unsound.

\begin{table}[t]

\small
\centering
\caption{Human labels of individual property assertions in LLM samples.}
\begin{tabular}{lcc}
\toprule
\textbf{Model} & \textbf{Sound} & \textbf{Unsound} \\ \midrule
Claude-3-Opus &  64 & 23 \\
Gemini-1.5-Pro & 23 & 5 \\
GPT-4 & 86 & 18 \\
 \textbf{Total} & 173 & 46 \\
 \bottomrule
\end{tabular}
\label{tab:soundness_label}
\end{table}

Table~\ref{tab:soundness_label} displays the results of the labeling. We note that the difference in the number of labeled assertions per model is due to the difference in validity and the number of property assertions in each test. We find that out of 219 labeled assertions, 173 (79\%) were sound. The soundness of assertions from each model were all similar, ranging from 74--83\%. To ensure the reliability of this manual labeling process, we computed the inter-rater agreement between two raters using Cohen's kappa coefficient ($\kappa$). The resulting $\kappa$ value of $0.862 \pm 0.133$ indicates a high level of agreement between the raters, suggesting that the labeling process was consistent.

We use our property assertion labels to determine soundness for each sample---if any assertions were unsound, the sample was labeled as unsound. We compared these labels to the labels resulting from our soundness metric and found that our metric has a precision of 100\% and recall of 97.14\%. There was only one false negative determined from the soundness metric due to the sample containing an unsound assertion that was not executed with an edge-case input in the 1,000 invocations.

When labeling property-based tests, we noticed a variety of soundness issues, ranging from hallucination of properties outside of the API documentation to assertions only failing due to very specific edge cases. The first test in Figure~\ref{fig:networkx-pbt-unsound} is an example in which a property assertion checks that when there is no cycle, the number of edges in a graph is less than the number of nodes. This property is true for undirected graphs, but not for directed graphs, which can be generated as inputs in this PBT. 

\begin{figure}[t]
\includegraphics[width=0.9\columnwidth]{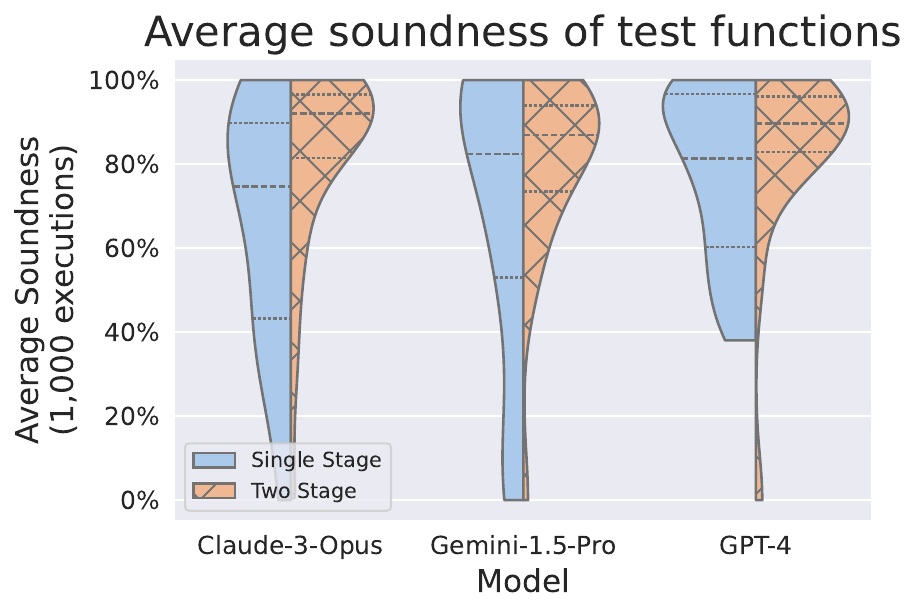}
\caption{Violin plot displaying soundness distribution of valid test functions, aggregated across all API methods (higher is better). Results for both single-stage and two-stage approaches are shown for each model. Overall, GPT-4 with two-stage prompting achieves the highest average soundness of % with an average of
87\% over all valid test functions.}
\label{fig:soundness}
\end{figure}

Another general pattern we noticed was that the property assertions held for \textit{most} of the generated inputs, but did not account for certain edge cases. This explains why many test functions have soundness of above 80\% in the distribution shown in Figure~\ref{fig:soundness}. An example of this is the second test in Figure~\ref{fig:networkx-pbt-unsound}. From the API docs, a property of the ISO calendar is that the ``first week of an ISO year is the first (Gregorian) calendar week of a year containing a Thursday.'' This PBT does not account for the case in which the first week of the Gregorian year does \textit{not} contain a Thursday, which occurs for a small percentage of generated inputs.

\begin{figure}[t]
    \begin{lstlisting}[style = python]
import ...

# Test for networkx.find_cycle
@given(st.data())
def test_find_cycle(data):
  (...) # Generator logic for graph G
  # verify the properties
  try:
    cycle = nx.find_cycle(G)
  except nx.exception.NetworkXNoCycle:
    # The graph doesn't have enough edges 
    # to form a cycle
    assert nx.number_of_edges(G) < num_nodes

# Test for date.isocalendar
@given(y=st.integers(min_value=1, max_value=9999),
   m=st.integers(min_value=1, max_value=12),
   d=st.integers(min_value=1, max_value=31))
def test_date_isocalendar(y, m, d):
  try:
    d = date(y, m, d)
    iso_year, iso_week, iso_weekday = d.isocalendar()

    assert iso_week >= 1 and iso_week <= 53
    assert iso_weekday >= 1 and iso_weekday <= 7
    assert iso_year == y or \
        (iso_week == 1 and iso_year == y + 1)
  except ValueError:
    pass
    \end{lstlisting}
    \caption{Example snippets from GPT-4 synthesized unsound property-based tests for \texttt{networkx.find\_cycle} and \texttt{datetime.date.isocalendar}. The second property assertion in \texttt{test\_find\_cycle} checks a general property of graphs that was not included in the API documentation that may be unsound when the graph is directed. The last assertion in \texttt{test\_date\_isocalendar} performs a comparison to the Gregorian year, but does not account for the case in which the first week of the Gregorian year does not contain a Thursday.}
    \label{fig:networkx-pbt-unsound}    
\end{figure}

\mybox{Through manual labeling on a set of LLM samples, we find that our soundness metric is aligned with human judgment, achieving \textbf{100\% precision and 97\% recall.} We observed that unsound assertions arose from hallucination of properties and failures to handle edge cases.}

\subsection{RQ3: Property Coverage}
\label{sec:eval:coverage}

\begin{table}[t]

\small
\centering
\caption{Property coverage of each model and prompting approach across all valid and sound LLM samples. Five property mutants were generated for each API, and mutants with runtime errors were filtered out.}
\begin{tabular}{llcc}
\toprule
\multirow{2}{*}{\textbf{Model}} & \multirow{2}{*}{\textbf{Approach}} & \textbf{Property Mutation} & \textbf{Total} \\ 
 & & \textbf{Score} & \textbf{Mutants} \\ 
\midrule
\multirow{2}{*}{Claude-3-Opus} & Single Stage & 58.02\% & 293 \\
       & Two Stage & 59.69\% & 258 \\ \midrule
\multirow{2}{*}{Gemini-1.5-Pro} & Single Stage & 43.57\% & 140 \\
       & Two Stage & 62.70\% & 244 \\ \midrule
\multirow{2}{*}{GPT-4} & Single Stage & 79.78\% & 89 \\
       & Two Stage & 51.88\% & 480  \\ \bottomrule
\end{tabular}
\label{tab:mutants}
\end{table}

% \begin{figure}[t]
% \centering
% \includegraphics[width=0.8\columnwidth]{figures/property-coverage.pdf}
% \caption{Bar plot displaying property coverage of all valid and sound test files, aggregated across all API methods. Results for both single-stage and two-stage approaches are shown for each model.}
% \label{fig:property-coverage}
% \end{figure}

RQ3 focuses on evaluating whether LLM-synthesized PBTs can detect property violations. Tests can be valid and sound while containing very simple assertions. \textit{Property coverage}, described in Section~\ref{sec:prop-coverage}, measures the ability of the property-based test to kill property mutants over all properties. This provides a measurement of the completeness of synthesized property-based test. Following the steps detailed in Section~\ref{sec:prop-coverage}, we prompt GPT-4 to extract five properties from the documentation for each API method and generate 5 property mutants for each property. In total, there are 200 properties we use to calculate overall property coverage.

We first measure \textit{property mutation score} over all valid and sound LLM. The tests must be sound to ensure that property mutants are only killed due to assertions checking the property mutant, rather than errors in the original test. We additionally filter any mutants that are killed due to validity errors to specifically measure the completeness of the property assertions. For each approach, we report the property mutation score as the average percent of property mutants killed over all samples. Table~\ref{tab:mutants} shows property mutation score for each approach as well as the total number of mutants that were executed. 

\begin{figure}[t]
\includegraphics[width=\columnwidth]{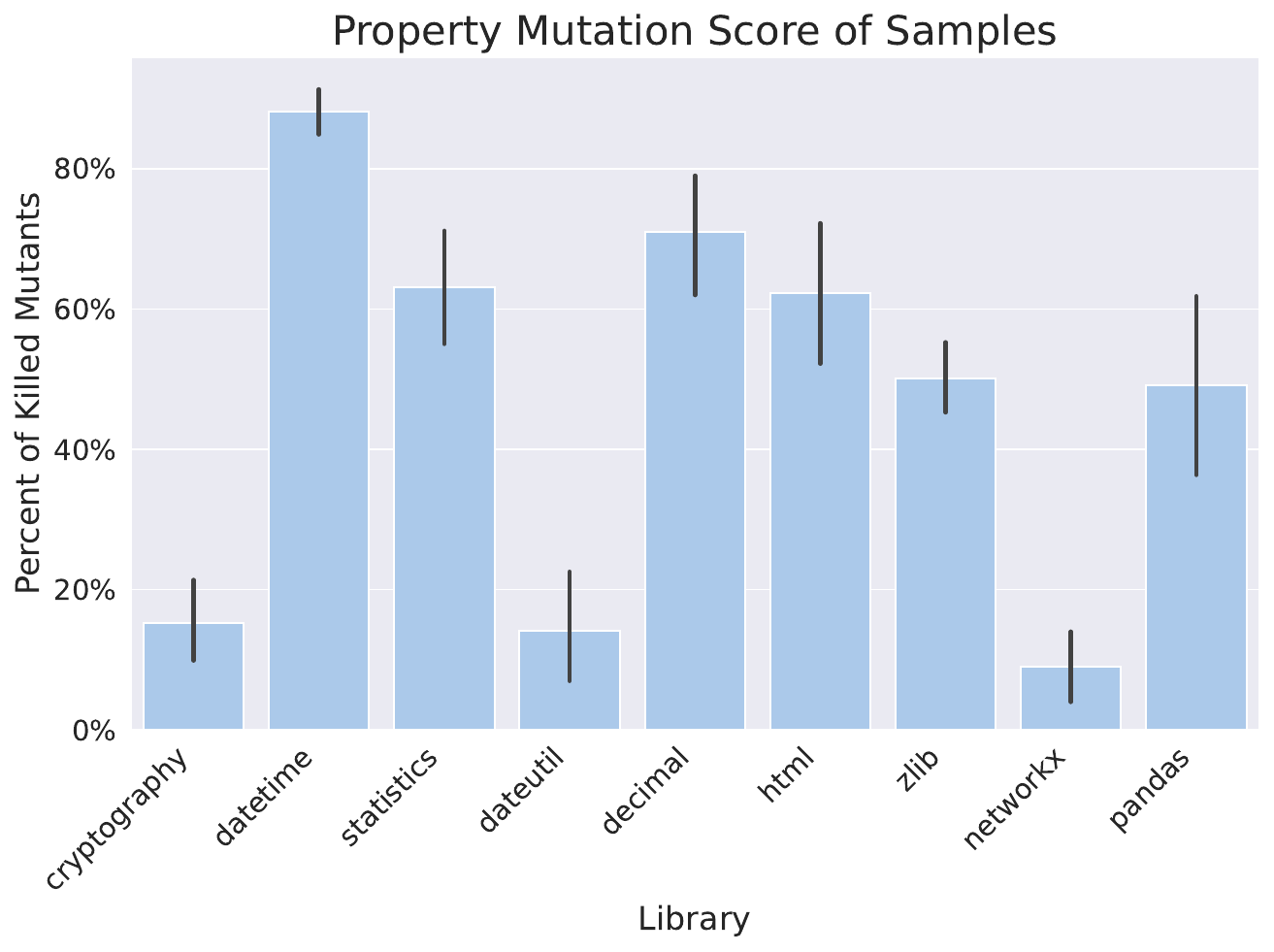}
\caption{Distribution of property mutation score over libraries across all valid and sound tests (higher is better). Generally, the tests achieve higher property mutation score from native Python libraries such as \texttt{statistics} and \texttt{datetime}. Property mutants from \texttt{networkx} require stronger assertions to detect.}
\label{fig:propcov}
\end{figure}

\begin{table}[t]

\small
\centering
\caption{Property coverage of LLM samples over all 40 API methods. In total, there were 200 properties across all API methods. If any of the five samples kills a property mutant, the property is covered.}
\begin{tabular}{llc}
\toprule
\multirow{2}{*}{\textbf{Model}} & \multirow{2}{*}{\textbf{Approach}} & \textbf{Property} \\ 
& & \textbf{Coverage} \\ \midrule
\multirow{2}{*}{Claude-3-Opus} & Single Stage & 9\% \\
       & Two Stage & 13\% \\ \midrule
\multirow{2}{*}{Gemini-1.5-Pro} & Single Stage & 5.5\% \\
       & Two Stage & 12\% \\ \midrule
\multirow{2}{*}{GPT-4} & Single Stage & 7\% \\
       & Two Stage & \textbf{20.50\%} \\ \bottomrule
\end{tabular}
\label{tab:propcov}
\end{table}

Overall, we observe that the valid and sound samples are able to kill 40--80\% of property mutants. These results vary across libraries, as shown in Figure~\ref{fig:propcov}. Samples achieve higher property coverage in native Python libraries such as \texttt{datetime} and \texttt{decimal}. \texttt{networkx} requires stronger assertions that are more difficult to synthesize due to the complexity of graph properties. 

We finally calculate the property coverage of all approaches on the entire set of 40 Python APIs. With five samples of an LLM for a synthesized PBT, how many properties can be covered? For a property to be covered, the sample must be valid, sound, and kill at least one property mutant. This measures the overall ability of the LLM to synthesize a PBT achieving all of our desired properties.  

Table~\ref{tab:propcov} shows the property coverage for each of our approaches. The best approach of two-stage prompting with GPT-4 is successfully able to synthesize property-based tests that are valid, sound, and cover 20.5\% of properties. We believe this is a promising result, as Proptest-AI is able to automate a significant portion of the property-based testing writing process and cover extracted docuemntation properties.

\mybox{The two-stage prompting approach with GPT-4 is successfully able to automatically synthesize PBTs covering \textbf{20.5\% of documented properties}.}

% \section{Discussion}
% The task of generating property-based tests using LLMs is challenging, and we have highlighted many different aspects of synthesizing and evaluating LLM samples. 
\section{Threats to Validity}
\subsection{Construct Validity}
Our measurements of validity and soundness are dependent on the executions of the property-based tests, which contain logic for random input generation. Additionally, due to the nondeterministic nature of LLMs, it is possible to get a range of samples that achieve varying degrees of performance. We aim to mitigate this nondeterminism by sampling from each LLM five times and executing the property-based tests 1,000 times, which is the standard for Hypothesis property-based tests. 
\subsection{Internal Validity}
One threat to internal validity comes from our execution-based definition of soundness. It is possible that an assertion error results from an actual bug rather than an unsound assertion. Our manual labeling process independently evaluated the soundness of property assertions, thus providing a stronger judgment. 

Another threat to internal validity is the existence of \textit{equivalent mutants}, which is a well studied problem in mutation testing~\cite{grun2009impact}. Many equivalent mutants arise due to the mutation at the source code level not propagating to the output of the program~\cite{just2014efficient}. Our formulation of property mutants applies the mutation \textit{directly} to the output of the API method rather than the source code, which mitigates this issue. However, in specific cases the output may remain equivalent after the mutation operator.

Finally, since we do not have access to the training set for any of the selected LLMs, we cannot confirm whether existing PBTs for our API methods are part of this set. However, we checked each of our selected Python library repositories and did not find any Hypothesis PBTs for any of the selected API methods. 

\subsection{External Validity}
We selected Python as a target language since LLMs have traditionally performed well with Python and Hypothesis is an extensive Python property-based testing library. We do not know of our conclusions will generalize to other programming languages or property-based testing libraries since this will impact the ability of the LLM to generate valid code. Another threat to external validity is the selection of our evaluation targets of Python API methods. We synthesized and evaluated property-based tests for a variety of native and popular Python libraries for which API documentation is readily available. Since these are established Python libraries, it is likely that the API documentation and source code exist in the training data of the LLMs. While we do not know of our conclusions will generalize to new libraries and APIs, prior work~\cite{zhou2022docprompting, su2024arks} has shown the effectiveness of providing documentation to LLMs to perform code generation for unknown APIs and languages.
\section{Related Work}
To the best of our knowledge, no work has yet attempted to use LLMs to automate the creation of property-based tests.  

LLMs have been used to aid in various testing tasks, including: generating tests cases for search-based unit test generation~\cite{lemieux2023codamosa}, co-generating code and unit-tests in an interactive manner~\cite{lahiri2022interactive}, and generating unit tests~\cite{bareiss2022pradeltest,schafer2023testpilot,rao2023cat,alshahwan2024automated}. TestPilot~\cite{schafer2023testpilot} uses fully automated prompt refinement when generating unit tests. In particular, when a test fails with an error, they enrich the prompt with various forms of information---including documentation snippets, the function signature, or the error encountered. 

In contrast, in TiCoder~\cite{lahiri2022interactive}, the goal is to synthesize the implementation of a function from natural language. The system concretizes this natural language spec by generating possible unit tests for the function being synthesized, and asking the user to approve the correctness of suggested input/output pairs.

% Our proposed evaluation methodology has links to problems faced in other realms of automated testing. % We discuss works touching on three broad aspects of automated testing. 
% \cl{I'm not sure I like this sectioning. I will wite and the rearrange.}

Regardless of whether they use machine learning, automated unit test generation techniques face some of the challenges we have outlined.
Randoop~\cite{pacheco2007randoop} generates only regression assertions. Regression assertions simply capture the (possibly buggy) behavior of the program under test at test generation time, and so, their \emph{soundness} with respect to the system specification is in question. The $\mu$Test~\cite{fraser2010mutest} approach uses mutation testing to reduce the number of assertions, and has been adopted in test suite generation systems such as Evosuite~\cite{fraser2011evosuite} and Pynguin~\cite{lukasczyk2022pynguin}. This addresses the \emph{completeness} of assertions, by keeping around only those assertions which can kill mutants. However, the problem of assertion \emph{soundness} remains.

The field of \emph{oracle generation} typically considers the problem of generating an \emph{oracle} (i.e., a set of property assertions) for a given test case that lacks assertions. 
% Given the soundness limitations of regression assertions, researchers have sought to extract oracles from other sources of information. 
One idea is to extract these from structured natural language information, such as JavaDoc comments.
From JavaDoc comments, TORADOCU~\cite{goffi2016toradocu} extracts exception oracles; MeMO~\cite{blasi2016Memo} extracts metamorphic relation oracles; and CallMeMaybe~\cite{blasi2023callmemaybe} extracts temporal property oracles. Of course, imprecision in the JavaDoc comments may lead to unsound assertions. However, the approaches have the potential to detect bugs in the implementation being analyzed, unlike regression oracles. As these tools use traditional NLP, the concerns of validity are less severe than for hallucination-prone LLM-based techniques.

Specialized deep learning techniques have also been proposed to generate oracles. ATLAS~\cite{watson2020learning} trains a Neural Machine Translation model on a dataset of (test case, oracle) pairs, and uses the trained model to generate oracles on new test cases. The NMT model predicts token sequences, avoiding some basic syntactic validity problems% \cl{we could probably say something more interesting in contrast with the technique... but I am losing energy in looking at the paper so I will stop here for now}
. TOGA~\cite{dinella2022toga} avoids the problem of validity by defining a \emph{grammar} of possible assertions, and having the deep learning model choose a production in this grammar. However, the problems of soundness and strength remain.

Unlike regular unit tests, which encode properties of a single input, the PBTs we generate encode properties over arbitrary generator-generated inputs. The problem of effectively searching the input space, so that the PBT shows a bug, is orthogonal to our current work. LLM-synthesized generators could be paired with coverage guidance~\cite{Lampropoulos19,padhye2019jqf}, validity-biased guidance~\cite{padhye2019zest}, reinforcement-learning guidance~\cite{reddy2020rlcheck}, or behavioral diversity guidance~\cite{nguyen2022bedivfuzz}, to produce more ``interesting'' test inputs. Fuzz4All~\cite{xia2024fuzz4all} %is a technique that even 
applies LLMs to guide and mutate input generation in fuzzing.

The problem of \emph{fuzz harness} or \emph{fuzz driver} generation bears similarity to our property generation problem~\cite{babic2019fudge,ispoglou2020fuzzgen}. In fuzz driver generation, the goal is to take unstructured byte data provided by the fuzz tester, and use it to exercise the program under test in a meaningful manner. The generated fuzz drivers resemble the ``combined'' PBT shown in Figure~\ref{fig:hypothesis-lists}, which consumes random \texttt{data} and uses it directly to construct inputs to exercise the API. As fuzz testing typically relies only on the crashing oracle (or on crashing oracles provided by instrumentation techniques such as ASAN), these works do not engage with the question of \emph{soundness} or \emph{property coverage} in the same manner we do. Nevertheless, the need for reasonable assertions emerges from a desire to reduce false positive bugs. The authors of UTopia~\cite{jeong2023utopia}, which extracts fuzz drivers from unit tests, note that some unit tests assertions (e.g., checking null pointers) must be preserved to maintain property validity. LLMs also have been applied for the task of fuzz driver generation~\cite{zhang2023understanding}. OSS-Fuzz~\cite{ossfuzz23} has a workflow to prompt LLMs for automated fuzz driver generation.

Finally, retrieval augmented code generation is a technique in natural that aims to improve the performance of code generation models by incorporating external knowledge from a large corpus of source code and related documentation. DocPrompting~\cite{zhou2022docprompting} uses retrieval to fetch relevant documentation pieces for a given natural-language-to-code task. ARKS~\cite{su2024arks} proposes \textit{active retrieval} for code generation, which evolves a ``knowledge soup'' integration many different forms of knowledge to prompt the LLM. We believe these strategies may provide methods of retrieving additional information that could be relevant to synthesizing property-based tests.

% The field of fuzz driver generation is a little closer in goal to our work. 
% %The problem of automatically generating properties bares some similarity to the problem of fuzz driver (or fuzz harness) generation. 
% In fuzz driver generation, the goal is typically to construct a sequence of API calls that can exercise some API under test in a low-false positive generators. Either this API consumes a byte sequence as input, or a methodology similar to the \texttt{data.draw} in Figure 2 is used. This is effectively the prefix of the property-based test, before the assertions. 

% Typically in fuzz driver generation, some default assertions (e.g., memory safety, are assumed). However, the problems of generator validity, generator diversity, and property validity have some analogues to the problems addressed in fuzz driver generation \cl{TODO: How so? add cites}.

\section{Data Availability}
We have included evaluation data in the anonymized repository at: \url{https://zenodo.org/doi/10.5281/zenodo.10967487}. This data contains all of the evaluation data for Proptest-AI, including synthesized property-based tests, property mutants, and measurements for validity, soundness, and property coverage. The API method documentation and prompt templates are also included.
\section{Conclusion}
In this paper, we explored and identified the unique challenges in using LLMs to synthesizing property-based tests. We characterized important properties of good property-based tests to propose an evaluation methodology that allows us to rigorously evaluate the LLM outputs. These properties include validity, soundness, and coverage of properties in the documentation. 

In our evaluation on 40 Python API methods and three state of the art language models, we find that our two-stage prompting approach with GPT-4 is able to produce valid and sound PBTs in 2.4 samples on average. This approach can automatically synthesize PBTs that cover 21\% of the documented properties. 

\begin{acks}
This research was supported in part by NSF Award CCF-2120955 and by the CyLab Future Enterprise Security Initiative.
\end{acks}

\bibliographystyle{IEEEtran}
\bibliography{IEEEabrv,references,codamosa-refs}

% Generated by IEEEtran.bst, version: 1.14 (2015/08/26)
\begin{thebibliography}{10}
\providecommand{\url}[1]{#1}
\csname url@samestyle\endcsname
\providecommand{\newblock}{\relax}
\providecommand{\bibinfo}[2]{#2}
\providecommand{\BIBentrySTDinterwordspacing}{\spaceskip=0pt\relax}
\providecommand{\BIBentryALTinterwordstretchfactor}{4}
\providecommand{\BIBentryALTinterwordspacing}{\spaceskip=\fontdimen2\font plus
\BIBentryALTinterwordstretchfactor\fontdimen3\font minus \fontdimen4\font\relax}
\providecommand{\BIBforeignlanguage}[2]{{%
\expandafter\ifx\csname l@#1\endcsname\relax
\typeout{** WARNING: IEEEtran.bst: No hyphenation pattern has been}%
\typeout{** loaded for the language `#1'. Using the pattern for}%
\typeout{** the default language instead.}%
\else
\language=\csname l@#1\endcsname
\fi
#2}}
\providecommand{\BIBdecl}{\relax}
\BIBdecl

\bibitem{claessen2000quickcheck}
K.~Claessen and J.~Hughes, ``Quickcheck: a lightweight tool for random testing of haskell programs,'' in \emph{Proceedings of the fifth ACM SIGPLAN international conference on Functional programming}, 2000, pp. 268--279.

\bibitem{arts2006testing}
T.~Arts, J.~Hughes, J.~Johansson, and U.~Wiger, ``Testing telecoms software with quviq quickcheck,'' in \emph{Proceedings of the 2006 ACM SIGPLAN Workshop on Erlang}, 2006, pp. 2--10.

\bibitem{arts2015testing}
T.~Arts, J.~Hughes, U.~Norell, and H.~Svensson, ``Testing autosar software with quickcheck,'' in \emph{2015 IEEE Eighth International Conference on Software Testing, Verification and Validation Workshops (ICSTW)}.\hskip 1em plus 0.5em minus 0.4em\relax IEEE, 2015, pp. 1--4.

\bibitem{hughes2016experiences}
J.~Hughes, ``Experiences with quickcheck: testing the hard stuff and staying sane,'' in \emph{A List of Successes That Can Change the World: Essays Dedicated to Philip Wadler on the Occasion of His 60th Birthday}.\hskip 1em plus 0.5em minus 0.4em\relax Springer, 2016, pp. 169--186.

\bibitem{hughes2016mysteries}
J.~Hughes, B.~C. Pierce, T.~Arts, and U.~Norell, ``Mysteries of dropbox: property-based testing of a distributed synchronization service,'' in \emph{2016 IEEE International Conference on Software Testing, Verification and Validation (ICST)}.\hskip 1em plus 0.5em minus 0.4em\relax IEEE, 2016, pp. 135--145.

\bibitem{padhye2019jqf}
R.~Padhye, C.~Lemieux, and K.~Sen, ``{JQF}: coverage-guided property-based testing in {Java},'' in \emph{Proceedings of the 28th ACM SIGSOFT International Symposium on Software Testing and Analysis}, 2019, pp. 398--401.

\bibitem{maciver2019hypothesis}
D.~R. MacIver, Z.~Hatfield-Dodds \emph{et~al.}, ``Hypothesis: A new approach to property-based testing,'' \emph{Journal of Open Source Software}, vol.~4, no.~43, p. 1891, 2019.

\bibitem{Lampropoulos19}
L.~Lampropoulos, M.~Hicks, and B.~C. Pierce, ``Coverage guided, property based testing,'' \emph{Proceedings of the ACM on Programming Languages}, vol.~3, no. OOPSLA, pp. 1--29, 2019.

\bibitem{DepsDev}
Google, ``{Open Source Insights},'' \url{https://deps.dev/}, retrieved April 27, 2023.

\bibitem{goldstein2024property}
H.~Goldstein, J.~W. Cutler, D.~Dickstein, B.~C. Pierce, and A.~Head, ``Property-based testing in practice,'' in \emph{2024 IEEE/ACM 46th International Conference on Software Engineering (ICSE)}.\hskip 1em plus 0.5em minus 0.4em\relax IEEE Computer Society, 2024, pp. 971--971.

\bibitem{chen2021evaluating}
M.~Chen, J.~Tworek, H.~Jun, Q.~Yuan, H.~P. d.~O. Pinto, J.~Kaplan, H.~Edwards, Y.~Burda, N.~Joseph, G.~Brockman \emph{et~al.}, ``{Evaluating Large Language Models Trained on Code},'' \emph{arXiv preprint arXiv:2107.03374}, 2021.

\bibitem{fried2022synthesis}
D.~Fried, A.~Aghajanyan, J.~Lin, S.~Wang, E.~Wallace, F.~Shi, R.~Zhong, W.-t. Yih, L.~Zettlemoyer, and M.~Lewis, ``Incoder: A generative model for code infilling and synthesis,'' \emph{arXiv preprint arXiv:2204.05999}, 2022.

\bibitem{bubeck2023sparks}
S.~Bubeck, V.~Chandrasekaran, R.~Eldan, J.~Gehrke, E.~Horvitz, E.~Kamar, P.~Lee, Y.~T. Lee, Y.~Li, S.~Lundberg, H.~Nori, H.~Palangi, M.~T. Ribeiro, and Y.~Zhang, ``Sparks of artificial general intelligence: Early experiments with gpt-4,'' 2023.

\bibitem{ouyang2022training}
L.~Ouyang, J.~Wu, X.~Jiang, D.~Almeida, C.~Wainwright, P.~Mishkin, C.~Zhang, S.~Agarwal, K.~Slama, A.~Ray \emph{et~al.}, ``Training language models to follow instructions with human feedback,'' \emph{Advances in Neural Information Processing Systems}, vol.~35, pp. 27\,730--27\,744, 2022.

\bibitem{openai2023gpt4}
OpenAI, ``Gpt-4 technical report,'' 2023.

\bibitem{lemieux2023codamosa}
C.~Lemieux, J.~P. Inala, S.~K. Lahiri, and S.~Sen, ``Codamosa: Escaping coverage plateaus in test generation with pre-trained large language models,'' in \emph{45th International Conference on Software Engineering, ser. ICSE}, 2023.

\bibitem{lahiri2022interactive}
\BIBentryALTinterwordspacing
S.~Lahiri, A.~Naik, G.~Sakkas, P.~Choudhury, C.~von Veh, M.~Musuvathi, J.~P. Inala, C.~Wang, and J.~Gao, ``Interactive code generation via test-driven user-intent formalization,'' arXiv, August 2022. [Online]. Available: \url{https://www.microsoft.com/en-us/research/publication/interactive-code-generation-via-test-driven-user-intent-formalization/}
\BIBentrySTDinterwordspacing

\bibitem{schafer2023testpilot}
M.~Schäfer, S.~Nadi, A.~Eghbali, and F.~Tip, ``Adaptive test generation using a large language model,'' 2023.

\bibitem{schafer2023empirical}
M.~Sch{\"a}fer, S.~Nadi, A.~Eghbali, and F.~Tip, ``An empirical evaluation of using large language models for automated unit test generation,'' \emph{IEEE Transactions on Software Engineering}, 2023.

\bibitem{zhang2023understanding}
C.~Zhang, M.~Bai, Y.~Zheng, Y.~Li, X.~Xie, Y.~Li, W.~Ma, L.~Sun, and Y.~Liu, ``Understanding large language model based fuzz driver generation,'' \emph{arXiv preprint arXiv:2307.12469}, 2023.

\bibitem{huang2024large}
L.~Huang, P.~Zhao, H.~Chen, and L.~Ma, ``Large language models based fuzzing techniques: A survey,'' \emph{arXiv preprint arXiv:2402.00350}, 2024.

\bibitem{ossfuzz23}
D.~Liu, J.~Metzman, and O.~Chang, ``{Open Source Insights},'' \url{https://security.googleblog.com/2023/08/ai-powered-fuzzing-breaking-bug-hunting.html}, 2023, retrieved February 27, 2024.

\bibitem{shoeybi2019megatron}
M.~Shoeybi, M.~Patwary, R.~Puri, P.~LeGresley, J.~Casper, and B.~Catanzaro, ``{Megatron-lm: Training multi-billion parameter language models using model parallelism},'' \emph{arXiv preprint arXiv:1909.08053}, 2019.

\bibitem{brown2020language}
T.~Brown, B.~Mann, N.~Ryder, M.~Subbiah, J.~D. Kaplan, P.~Dhariwal, A.~Neelakantan, P.~Shyam, G.~Sastry, A.~Askell \emph{et~al.}, ``Language models are few-shot learners,'' \emph{Advances in neural information processing systems}, vol.~33, pp. 1877--1901, 2020.

\bibitem{chowdhery2022palm}
A.~Chowdhery, S.~Narang, J.~Devlin, M.~Bosma, G.~Mishra, A.~Roberts, P.~Barham, H.~W. Chung, C.~Sutton, S.~Gehrmann \emph{et~al.}, ``{Palm: Scaling language modeling with pathways},'' \emph{arXiv preprint arXiv:2204.02311}, 2022.

\bibitem{thoppilan2022lamda}
R.~Thoppilan, D.~De~Freitas, J.~Hall, N.~Shazeer, A.~Kulshreshtha, H.-T. Cheng, A.~Jin, T.~Bos, L.~Baker, Y.~Du \emph{et~al.}, ``{Lamda: Language models for dialog applications},'' \emph{arXiv preprint arXiv:2201.08239}, 2022.

\bibitem{reid2024gemini}
M.~Reid, N.~Savinov, D.~Teplyashin, D.~Lepikhin, T.~Lillicrap, J.-b. Alayrac, R.~Soricut, A.~Lazaridou, O.~Firat, J.~Schrittwieser \emph{et~al.}, ``Gemini 1.5: Unlocking multimodal understanding across millions of tokens of context,'' \emph{arXiv preprint arXiv:2403.05530}, 2024.

\bibitem{claude3}
Anthropic, ``{Introducing the next generation of Claude},'' \url{https://www.anthropic.com/news/claude-3-family}, retrieved April 1, 2024.

\bibitem{liu2023pre}
P.~Liu, W.~Yuan, J.~Fu, Z.~Jiang, H.~Hayashi, and G.~Neubig, ``Pre-train, prompt, and predict: A systematic survey of prompting methods in natural language processing,'' \emph{ACM Computing Surveys}, vol.~55, no.~9, pp. 1--35, 2023.

\bibitem{xu2022polycoder}
F.~F. Xu, U.~Alon, G.~Neubig, and V.~J. Hellendoorn, ``A systematic evaluation of large language models of code,'' in \emph{Proceedings of the 6th ACM SIGPLAN International Symposium on Machine Programming}, 2022, pp. 1--10.

\bibitem{li2023starcoder}
\BIBentryALTinterwordspacing
R.~Li, L.~Ben~Allal, Y.~Zi, N.~Muennighoff, D.~Kocetkov, C.~Mou, M.~Marone, C.~Akiki, J.~Li, J.~Chim, Q.~Liu, E.~Zheltonozhskii, T.~Y. Zhuo, T.~Wang, O.~Dehaene, M.~Davaadorj, J.~Lamy-Poirier, J.~Monteiro, O.~Shliazhko, N.~Gontier, N.~Meade, A.~Randy, M.-H. Yee, L.~K. Umapathi, J.~Zhu, B.~Lipkin, M.~Oblokulov, Z.~Wang, R.~Murthy, J.~Stillerman, S.~S. Patel, D.~Abulkhanov, M.~Zocca, M.~Dey, Z.~Zhang, N.~Fahmy, U.~Bhattacharyya, S.~Gunasekar, W.~Yu, S.~Singh, S.~Luccioni, P.~Villegas, M.~Kunakov, F.~Zhdanov, M.~Romero, T.~Lee, N.~Timor, J.~Ding, C.~Schlesinger, H.~Schoelkopf, J.~Ebert, T.~Dao, M.~Mishra, A.~Gu, J.~Robinson, C.~J. Anderson, B.~Dolan-Gavitt, D.~Contractor, S.~Reddy, D.~Fried, D.~Bahdanau, Y.~Jernite, C.~M. Ferrandis, S.~Hughes, T.~Wolf, A.~Guha, L.~von Werra, and H.~de~Vries, ``{STARCODER: May the Source be With You!}'' 2023. [Online]. Available: \url{https://drive.google.com/file/d/1cN-b9GnWtHzQRoE7M7gAEyivY0kl4BYs/view}
\BIBentrySTDinterwordspacing

\bibitem{roziere2023code}
B.~Roziere, J.~Gehring, F.~Gloeckle, S.~Sootla, I.~Gat, X.~E. Tan, Y.~Adi, J.~Liu, T.~Remez, J.~Rapin \emph{et~al.}, ``Code llama: Open foundation models for code,'' \emph{arXiv preprint arXiv:2308.12950}, 2023.

\bibitem{guo2024deepseek}
D.~Guo, Q.~Zhu, D.~Yang, Z.~Xie, K.~Dong, W.~Zhang, G.~Chen, X.~Bi, Y.~Wu, Y.~Li \emph{et~al.}, ``Deepseek-coder: When the large language model meets programming--the rise of code intelligence,'' \emph{arXiv preprint arXiv:2401.14196}, 2024.

\bibitem{austin2021synthesis}
J.~Austin, A.~Odena, M.~Nye, M.~Bosma, H.~Michalewski, D.~Dohan, E.~Jiang, C.~Cai, M.~Terry, Q.~Le \emph{et~al.}, ``{Program synthesis with large language models},'' \emph{arXiv preprint arXiv:2108.07732}, 2021.

\bibitem{prenner2021progrepair}
J.~A. Prenner and R.~Robbes, ``{Automatic Program Repair with OpenAI's Codex: Evaluating QuixBugs},'' \emph{arXiv preprint arXiv:2111.03922}, 2021.

\bibitem{pearce2021fixbugs}
H.~Pearce, B.~Tan, B.~Ahmad, R.~Karri, and B.~Dolan-Gavitt, ``{Can OpenAI Codex and Other Large Language Models Help Us Fix Security Bugs?}'' \emph{arXiv preprint arXiv:2112.02125}, 2021.

\bibitem{pearce2023vulrepair}
\BIBentryALTinterwordspacing
------, ``Examining zero-shot vulnerability repair with large language models,'' in \emph{2023 2023 IEEE Symposium on Security and Privacy (SP) (SP)}.\hskip 1em plus 0.5em minus 0.4em\relax Los Alamitos, CA, USA: IEEE Computer Society, may 2023, pp. 1--18. [Online]. Available: \url{https://doi.ieeecomputersociety.org/10.1109/SP46215.2023.00001}
\BIBentrySTDinterwordspacing

\bibitem{sarsa2022explanations}
S.~Sarsa, P.~Denny, A.~Hellas, and J.~Leinonen, ``{Automatic Generation of Programming Exercises and Code Explanations Using Large Language Models},'' in \emph{Proceedings of the 2022 ACM Conference on International Computing Education Research V. 1}, 2022, pp. 27--43.

\bibitem{kojima2022large}
T.~Kojima, S.~S. Gu, M.~Reid, Y.~Matsuo, and Y.~Iwasawa, ``Large language models are zero-shot reasoners,'' \emph{Advances in neural information processing systems}, vol.~35, pp. 22\,199--22\,213, 2022.

\bibitem{goodenough1975toward}
J.~B. Goodenough and S.~L. Gerhart, ``Toward a theory of test data selection,'' in \emph{Proceedings of the international conference on Reliable software}, 1975, pp. 493--510.

\bibitem{frankl1998further}
P.~G. Frankl and O.~Iakounenko, ``Further empirical studies of test effectiveness,'' in \emph{Proceedings of the 6th ACM SIGSOFT international symposium on Foundations of software engineering}, 1998, pp. 153--162.

\bibitem{fraser2014large}
G.~Fraser and A.~Arcuri, ``A large-scale evaluation of automated unit test generation using evosuite,'' \emph{ACM Transactions on Software Engineering and Methodology (TOSEM)}, vol.~24, no.~2, pp. 1--42, 2014.

\bibitem{shamshiri2015automatically}
S.~Shamshiri, R.~Just, J.~M. Rojas, G.~Fraser, P.~McMinn, and A.~Arcuri, ``Do automatically generated unit tests find real faults? an empirical study of effectiveness and challenges (t),'' in \emph{2015 30th IEEE/ACM International Conference on Automated Software Engineering (ASE)}.\hskip 1em plus 0.5em minus 0.4em\relax IEEE, 2015, pp. 201--211.

\bibitem{fan2023large}
A.~Fan, B.~Gokkaya, M.~Harman, M.~Lyubarskiy, S.~Sengupta, S.~Yoo, and J.~M. Zhang, ``Large language models for software engineering: Survey and open problems,'' \emph{arXiv preprint arXiv:2310.03533}, 2023.

\bibitem{DeMillo78}
R.~DeMillo, R.~Lipton, and F.~Sayward, ``Hints on test data selection: Help for the practicing programmer,'' \emph{Computer}, vol.~11, no.~4, pp. 34--41, 1978.

\bibitem{ibrahimzada2023automated}
A.~R. Ibrahimzada, Y.~Chen, R.~Rong, and R.~Jabbarvand, ``Automated bug generation in the era of large language models,'' \emph{arXiv preprint arXiv:2310.02407}, 2023.

\bibitem{gargcoupling}
A.~Garg, R.~Degiovanni, M.~Papadakis, and Y.~Le~Traon, ``On the coupling between vulnerabilities and llm-generated mutants: A study on vul4j dataset.''

\bibitem{achiam2023gpt}
J.~Achiam, S.~Adler, S.~Agarwal, L.~Ahmad, I.~Akkaya, F.~L. Aleman, D.~Almeida, J.~Altenschmidt, S.~Altman, S.~Anadkat \emph{et~al.}, ``Gpt-4 technical report,'' \emph{arXiv preprint arXiv:2303.08774}, 2023.

\bibitem{grun2009impact}
B.~J. Gr{\"u}n, D.~Schuler, and A.~Zeller, ``The impact of equivalent mutants,'' in \emph{2009 International Conference on Software Testing, Verification, and Validation Workshops}.\hskip 1em plus 0.5em minus 0.4em\relax IEEE, 2009, pp. 192--199.

\bibitem{just2014efficient}
R.~Just, M.~D. Ernst, and G.~Fraser, ``Efficient mutation analysis by propagating and partitioning infected execution states,'' in \emph{Proceedings of the 2014 international symposium on software testing and analysis}, 2014, pp. 315--326.

\bibitem{zhou2022docprompting}
S.~Zhou, U.~Alon, F.~F. Xu, Z.~Wang, Z.~Jiang, and G.~Neubig, ``Docprompting: Generating code by retrieving the docs,'' \emph{arXiv preprint arXiv:2207.05987}, 2022.

\bibitem{su2024arks}
H.~Su, S.~Jiang, Y.~Lai, H.~Wu, B.~Shi, C.~Liu, Q.~Liu, and T.~Yu, ``Arks: Active retrieval in knowledge soup for code generation,'' \emph{arXiv preprint arXiv:2402.12317}, 2024.

\bibitem{bareiss2022pradeltest}
P.~Barei{\ss}, B.~Souza, M.~d'Amorim, and M.~Pradel, ``Code generation tools (almost) for free? a study of few-shot, pre-trained language models on code,'' \emph{arXiv preprint arXiv:2206.01335}, 2022.

\bibitem{rao2023cat}
N.~Rao, K.~Jain, U.~Alon, C.~Le~Goues, and V.~J. Hellendoorn, ``Cat-lm training language models on aligned code and tests,'' in \emph{2023 38th IEEE/ACM International Conference on Automated Software Engineering (ASE)}.\hskip 1em plus 0.5em minus 0.4em\relax IEEE, 2023, pp. 409--420.

\bibitem{alshahwan2024automated}
N.~Alshahwan, J.~Chheda, A.~Finegenova, B.~Gokkaya, M.~Harman, I.~Harper, A.~Marginean, S.~Sengupta, and E.~Wang, ``Automated unit test improvement using large language models at meta,'' \emph{arXiv preprint arXiv:2402.09171}, 2024.

\bibitem{pacheco2007randoop}
C.~Pacheco and M.~D. Ernst, ``Randoop: feedback-directed random testing for {Java},'' in \emph{Companion to the 22nd ACM SIGPLAN conference on Object-oriented programming systems and applications companion}, 2007, pp. 815--816.

\bibitem{fraser2010mutest}
\BIBentryALTinterwordspacing
G.~Fraser and A.~Zeller, ``Mutation-driven generation of unit tests and oracles,'' in \emph{Proceedings of the 19th International Symposium on Software Testing and Analysis}, ser. ISSTA '10.\hskip 1em plus 0.5em minus 0.4em\relax New York, NY, USA: Association for Computing Machinery, 2010, p. 147–158. [Online]. Available: \url{https://doi.org/10.1145/1831708.1831728}
\BIBentrySTDinterwordspacing

\bibitem{fraser2011evosuite}
G.~Fraser and A.~Arcuri, ``Evosuite: automatic test suite generation for object-oriented software,'' in \emph{Proceedings of the 19th ACM SIGSOFT symposium and the 13th European conference on Foundations of software engineering}, 2011, pp. 416--419.

\bibitem{lukasczyk2022pynguin}
S.~Lukasczyk and G.~Fraser, ``Pynguin: Automated unit test generation for python,'' in \emph{Proceedings of the ACM/IEEE 44th International Conference on Software Engineering: Companion Proceedings}, 2022, pp. 168--172.

\bibitem{goffi2016toradocu}
\BIBentryALTinterwordspacing
A.~Goffi, A.~Gorla, M.~D. Ernst, and M.~Pezz\`{e}, ``Automatic generation of oracles for exceptional behaviors,'' in \emph{Proceedings of the 25th International Symposium on Software Testing and Analysis}, ser. ISSTA 2016.\hskip 1em plus 0.5em minus 0.4em\relax New York, NY, USA: Association for Computing Machinery, 2016, p. 213–224. [Online]. Available: \url{https://doi.org/10.1145/2931037.2931061}
\BIBentrySTDinterwordspacing

\bibitem{blasi2016Memo}
\BIBentryALTinterwordspacing
A.~Blasi, A.~Gorla, M.~D. Ernst, M.~Pezzè, and A.~Carzaniga, ``{MeMo: Automatically identifying metamorphic relations in Javadoc comments for test automation},'' \emph{Journal of Systems and Software}, vol. 181, p. 111041, 2021. [Online]. Available: \url{https://www.sciencedirect.com/science/article/pii/S0164121221001382}
\BIBentrySTDinterwordspacing

\bibitem{blasi2023callmemaybe}
\BIBentryALTinterwordspacing
A.~Blasi, A.~Gorla, M.~D. Ernst, and M.~Pezz\`{e}, ``{Call Me Maybe: Using NLP to Automatically Generate Unit Test Cases Respecting Temporal Constraints},'' in \emph{Proceedings of the 37th IEEE/ACM International Conference on Automated Software Engineering}, ser. ASE '22.\hskip 1em plus 0.5em minus 0.4em\relax New York, NY, USA: Association for Computing Machinery, 2023. [Online]. Available: \url{https://doi.org/10.1145/3551349.3556961}
\BIBentrySTDinterwordspacing

\bibitem{watson2020learning}
C.~Watson, M.~Tufano, K.~Moran, G.~Bavota, and D.~Poshyvanyk, ``On learning meaningful assert statements for unit test cases,'' in \emph{Proceedings of the ACM/IEEE 42nd International Conference on Software Engineering}, 2020, pp. 1398--1409.

\bibitem{dinella2022toga}
E.~Dinella, G.~Ryan, T.~Mytkowicz, and S.~K. Lahiri, ``Toga: a neural method for test oracle generation,'' in \emph{Proceedings of the 44th International Conference on Software Engineering}, 2022, pp. 2130--2141.

\bibitem{padhye2019zest}
R.~Padhye, C.~Lemieux, K.~Sen, M.~Papadakis, and Y.~Le~Traon, ``{Semantic Fuzzing with Zest},'' in \emph{Proceedings of the 28th ACM SIGSOFT International Symposium on Software Testing and Analysis}, 2019, pp. 329--340.

\bibitem{reddy2020rlcheck}
S.~Reddy, C.~Lemieux, R.~Padhye, and K.~Sen, ``Quickly generating diverse valid test inputs with reinforcement learning,'' in \emph{Proceedings of the ACM/IEEE 42nd International Conference on Software Engineering}, 2020, pp. 1410--1421.

\bibitem{nguyen2022bedivfuzz}
H.~L. Nguyen and L.~Grunske, ``Bedivfuzz: integrating behavioral diversity into generator-based fuzzing,'' in \emph{Proceedings of the 44th International Conference on Software Engineering}, 2022, pp. 249--261.

\bibitem{xia2024fuzz4all}
C.~S. Xia, M.~Paltenghi, J.~Le~Tian, M.~Pradel, and L.~Zhang, ``Fuzz4all: Universal fuzzing with large language models,'' \emph{Proc. IEEE/ACM ICSE}, 2024.

\bibitem{babic2019fudge}
D.~Babi{\'c}, S.~Bucur, Y.~Chen, F.~Ivan{\v{c}}i{\'c}, T.~King, M.~Kusano, C.~Lemieux, L.~Szekeres, and W.~Wang, ``Fudge: fuzz driver generation at scale,'' in \emph{Proceedings of the 2019 27th ACM Joint Meeting on European Software Engineering Conference and Symposium on the Foundations of Software Engineering}, 2019, pp. 975--985.

\bibitem{ispoglou2020fuzzgen}
K.~K. Ispoglou, D.~Austin, V.~Mohan, and M.~Payer, ``Fuzzgen: Automatic fuzzer generation,'' in \emph{Proceedings of the 29th USENIX Conference on Security Symposium}, 2020, pp. 2271--2287.

\bibitem{jeong2023utopia}
\BIBentryALTinterwordspacing
B.~Jeong, J.~Jang, H.~Yi, J.~Moon, J.~Kim, I.~Jeon, T.~Kim, W.~Shim, and Y.~Hwang, ``Utopia: Automatic generation of fuzz driver using unit tests,'' in \emph{2023 2023 IEEE Symposium on Security and Privacy (SP) (SP)}.\hskip 1em plus 0.5em minus 0.4em\relax Los Alamitos, CA, USA: IEEE Computer Society, may 2023, pp. 746--762. [Online]. Available: \url{https://doi.ieeecomputersociety.org/10.1109/SP46215.2023.00043}
\BIBentrySTDinterwordspacing

\end{thebibliography}

\end{document}